\documentclass[draft]{agujournal2019}

\usepackage{graphicx} % Required for inserting images

\usepackage{url} %this package should fix any errors with URLs in refs.
\usepackage{lineno}
\usepackage{amsmath}
\usepackage{soul}
\usepackage[inline]{trackchanges} %for better track changes. finalnew option will compile document with changes incorporated.
\usepackage{natbib}
\journalname{Biogeosciences}

\begin{document}

\title{Deep learning model emulators for marine biogeochemistry forecasting from days to decades}
\authors{Jozef Sk\'akala$^{1,2}$, Ieuan Higgs$^{3}$, David Moffat$^{1,2}$ \\
\emph{$^{1}$Plymouth Marine Laboratory, Plymouth, UK \\
    $^{2}$National Centre for Earth Observation, Plymouth, UK \\
    $^{3}$Department of Computer Science, University of Reading, Reading, UK}}

\correspondingauthor{Jozef Sk\'akala}{jos@pml.ac.uk}

\date{May 2026}

%\maketitle

\begin{abstract}
There has been a major surge of activity in the development of deep-learning emulators for Earth System Models. These emulators offer the potential to substantially reduce computational costs, when used to run short-range, seasonal, and climate-scale predictions. When trained on reanalysis data, they may even outperform conventional numerical model forecasts. Marine biogeochemistry has so far lagged somewhat behind these developments, although a rapid expansion of activity in this area is already happening. We have used a simplified one-dimensional water column model, coupling a physical ocean model with a high complexity marine biogeochemistry model (European Regional Seas Ecosystem Model, ERSEM), and have demonstrated that ERSEM can be successfully emulated using deep learning techniques. We explored two key emulator architectures, Long Short-Term Memory (LSTM) Neural Networks that emulate a selected subset of ERSEM variables at daily time-step resolution, and physics-informed 1D Convolutional Neural Networks (CNN) that emulate the full pelagic ERSEM system throughout the entire water column.
We show that by using ocean physics simulator inputs, these emulators remain largely stable over multi-decadal timescales and are highly skilful in reproducing ERSEM simulations in both decadal climate projections and short-range (10-day) forecasting applications. The former includes the emulator's ability to accurately predict the timing of phytoplankton Spring blooms several years in advance. Furthermore, we show that, when trained on reanalysis data, the emulators can outperform ERSEM forecast skill score for several key variables, including phytoplankton and zooplankton, by 50–60\%. If similar performance can be achieved in three-dimensional regional applications, the emulators could deliver substantially higher-quality predictions at a fraction of the computational cost.
%The former includes good skill of the emulator to predict phytoplankton Spring bloom timing years ahead. %Furthermore, we demonstrate that when trained on reanalysis data, the emulators can outperform ERSEM forecast skill of some key variables, such as phytoplankton and zooplankton, by 50-60\%. If similar performance remained in three-dimensional regional application, the emulators would deliver much higher-quality predictions at a fraction of the computational cost. 
We also apply novel explainability techniques, which can help provide insights into emulated system's emergent behaviour and feed important information to marine biogeochemistry model builders. The emulator performance is evaluated using a range of metrics, including the ability to reproduce daily anomalies and extreme events. We anticipate that these emulator techniques will have wide applicability in the future, including operational forecasting and in marine autonomous systems. We conclude by discussing key challenges and opportunities for further development.
    
\end{abstract}

\section{Introduction}

%\begin{itemize}
%\item Discuss that emulators are extremely timely topic in forecasting and what is their purpose (reduction of comp cost and potentially improved forecasts by learning from reanalysis rather than the free run)
%\item in weather prediction, pioneering results have been achieved by ECMWF, other regional forecasters such as NIWA in New Zealand.
%\item In ocean physics - a very recent surge of results, some challenges in interpretation (two-way coupling with atmosphere).
%\item In marine BGC, things are a little behind although first examples appeared (Smith et al. 2026) - replacing marine BGC component provides paradoxically a clearer case for emulators than the physics (lesser two-way interaction problem).
%\item the purpose of this paper is to demonstrate two prototypes of emulators for a highly complex BGC model and explore their different uses. The most straightforward option to explore these questions is the 1D watercolumn modelling.
%\end{itemize}

There has been a rapid growth in the development and application of Artificial Intelligence (AI)-based forecasting systems. In recent years, these systems have been developed by international operational centres for applications such as (i) short- and medium-range atmospheric forecasting \citep[e.g.,][]{pathak2022fourcastnet, lam2023learning, bi2023accurate, li2024generative, bonev2025fourcastnet}, including the AI Forecasting System (AIFS) of the European Centre for Medium-Range Weather Forecasts \citep[ECMWF,][]{lang2024aifs}; (ii) regional climate downscaling \citep{rampal2025reliable}; and (iii) forecasting of extreme atmospheric events \citep{jimenez2025ai, camps2025artificial}. More recently, substantial effort has also been devoted to the development of deep learning emulators of physical ocean models for ocean forecasting \citep{xiong2023ai, guo2024orca, subel2024building, holmberg2024regional, dheeshjith2025samudra, gray2025long}, including coupled atmosphere–ocean emulators \citep{duncan2025samudrace}. 
One key motivation for deep learning–based forecasting systems is their ability to substantially reduce computational expense. For example, atmospheric emulator FourCastNet achieved an O$(10^{5})$ speed-up relative to traditional numerical weather prediction systems \citep{pathak2022fourcastnet}. While training such models can be computationally intensive, their operational cost is typically very low. Moreover, when trained on reanalysis products rather than free-running simulations, deep learning emulators may improve forecast quality by learning from observationally constrained representations of the Earth system \citep{lam2023learning}.
%One key motivation for these deep learning-based forecasting systems is to substantially reduce computational expense (e.g. a O($10^{5}$) speed-up was reported for FourCastNet in \citet{pathak2022fourcastnet}), as deep learning models are typically inexpensive to run (although potentially expensive to train). In the same time, the emulators may potentially improve forecast quality when trained on reanalysis data rather than free-running model simulations \citep{lam2023learning}. 

Research on AI-based forecasting systems for marine biogeochemistry has progressed more slowly. Previous work has primarily focused on emulating specific aspects of model behaviour, including parameter estimation \citep[e.g.,][]{mattern2012estimating, schartau2017reviews}, uncertainty propagation \citep{mattern2013sensitivity}, and the inference of individual biogeochemical variables from model outputs \citep[e.g.,][]{skakala2023future, skakala2026estimating, le2026equation}. However, an example of a dynamical AI emulator for marine biogeochemical forecasting was recently presented by \cite{smith2026identifying}. In that study, several key variables, including chlorophyll-$a$ and nutrients averaged over the upper ocean layer, were forecast at short- to medium-range timescales (up to 16 days), demonstrating the emulator's ability to reproduce process-based model forecasts in the Black Sea environment.

Marine biogeochemistry models typically contain much larger number of prognostic state variables and interacting processes than the ocean physics models. The more complex of these models can easily include tens of pelagic variables, as well as additional variables for benthic fauna \citep[e.g.,][]{fennel2022ocean}.
The large complexity of marine biogeochemistry models incurs significant computational cost, e.g. depending on their complexity, adding marine biogeochemistry model typically increases the cost of a marine physics model by a factor of 2-10 \citep{kwiatkowski2014imarnet}. The high cost makes the emulation of marine biogeochemistry models particularly desirable, but the high complexity makes it also potentially challenging. 

Marine biogeochemistry models encompass a wide range of highly nonlinear, often incompletely understood, processes that couple degrees of freedom both within and across lower-trophic-level food webs, non-living organic matter pools, and inorganic constituents. The quality of short-range and long-term forecasts are affected by the phenomenological nature of state-of-the-art marine biogeochemical models, their large number of poorly constrained parameters \citep{schartau2017reviews}, and the strongly nonlinear and non-Gaussian characteristics of the underlying dynamics \citep[e.g.,][]{fennel2019advancing}. For example, in temperate shelf seas around the UK, short-range forecasts of key biogeochemical variables such as chlorophyll-$a$ exhibit large, seasonally varying biases \citep[e.g.,][]{skakala2018assimilation, fowler2023validating, banerjee2026assimilation}. In this context, deep learning emulators trained on reanalysis products may be particularly valuable, as they can implicitly learn and correct systematic biases present in process-based forecasting systems. 

Furthermore, although marine physics and biogeochemistry are coupled in reality, the feedback from biogeochemistry to physical ocean processes is often comparatively weak and can frequently be represented through external forcing or parameterisations \citep{fennel2022ocean}. As a result, ocean circulation models are commonly run in a stand-alone configuration and, when biogeochemistry is included, it is typically coupled only in one direction: physical fields drive biogeochemical processes, while the biogeochemical state has little or no influence on the simulated physics \citep{ford2018marine}. This characteristic is advantageous for AI-based biogeochemical emulation. Because the emulator typically needs to rely on externally provided atmospheric and physical-ocean inputs, the one-way coupling between ocean physics and biogeochemistry means it can be applied consistently across a wide range of forecasting and projection timescales. This contrasts with the emulation of ocean physics from atmospheric forcing, where physically consistent representation of ocean–atmosphere feedbacks is essential and therefore often requires fully coupled emulator frameworks, such as that proposed by \citet{duncan2025samudrace}.
 
A further, emerging advantage of using deep learning approaches to emulate biogeochemistry models lies in explainable AI (XAI) methods, or AI-enabled diagnostics \citep{gevaert2022explainable, bacsaugaouglu2022review}. These can be applied to a trained emulator to deepen understanding of emergent process-level model behaviours and dependencies.
For such systems, XAI approaches can efficiently reveal relationships that are difficult to capture using traditional multivariate statistical techniques, which often rely on linear assumptions, perturbation methods, or low-order interactions \citep{bacsaugaouglu2022review}.
In this way, we can build AI-enhanced understanding of biogeochemistry models, benefiting the scientific community of model-builders. This could foster a symbiotic relationship between traditional model development and AI-based emulation: XAI diagnostics may reveal that an emulator relies on a process representation or parameter sensitivity not anticipated by the original model structure, prompting model-builders to revisit and refine the underlying mechanistic formulation; in turn, improvements to the process-based model provide better training data and structure for the next generation of emulators, sharpening their predictive accuracy and the diagnostics drawn from them.

In this work, we explore a range of approaches (e.g., short-range/decadal, model architectures, trained by reanalysis/free run) for emulating a complex UK regional marine biogeochemical model using deep learning. These are mainly developed along two types of deep learning emulators for the dynamical system time-evolution: (i) a physics-informed 1D Convolutional Neural Network \citep[CNN,][]{lecun1995cnn} based on a ResNet architecture~\citep{he2016deep}, emulating a daily time-step operator of the full pelagic system represented by the biogeochemistry model across the whole water column, and (ii) a Long-Short Term Memory (LSTM) neural network \citep{hochreiter1997long} representing a daily time-step operator for a selected number of variables mostly at the ocean surface. We assess the forecast skill of the proposed emulators across a range of timescales, from short- to medium-range forecasts (up to 10 days) to decadal prediction horizons. We also discuss emulator stability and how this relates to the two different architectures, and examine whether training on a biogeochemical reanalysis can improve forecast performance relative to the process-based simulator. We deploy deep ensemble techniques to assess the sensitivity of the emulator forecast to the training configuration, assess the impact of imperfect inputs on the emulator stability, and provide explainability of the results. Finally, we explicitly demonstrate how XAI methods can provide insights not only into the behaviour of the deep-learning emulators but also, potentially, into the underlying process-based simulator, for example by elucidating the key drivers of Spring bloom formation and hypoxic conditions. 

To make the problem tractable, we focus on a simplified one-dimensional water column configuration of a regional UK marine biogeochemical model at a coastal site in the western English Channel. Although one-dimensional model is computationally cheap and does not need to be emulated, it has often proven to be a valuable test-bed for ideas which can then be applied in the three-dimensional regional model. Indeed one-dimensional models are commonly used for computationally expensive tasks, examples include developing methods to estimate time-varying biogeochemical model parameters \citep{schartau2003simultaneous, simon2012gaussian}, to assess the impact of atmospheric forcing resolution on ecosystem dynamics \citep{powley2020sensitivity}, assess model observability and controllability \citep{ciavatta2025control}, evaluate the impact of model complexity on model skill and portability \citep{friedrichs2007assessment}, investigate new hybrid machine learning - data assimilation approaches \citep{higgs2025hybrid}, or to evaluate the impact of sampling frequency and observational uncertainty on multi-platform data assimilation \citep{cossarini2026improving}. Although the one-dimensional model does not account for regional exchange through horizontal advection, or for important external forcings such as river discharge, it retains the full complexity of the local biogeochemical dynamics and their response to physical mixing, salinity variability, and air–sea heat fluxes. We recognize that the additional complexity of the three-dimensional dynamics will likely require additional deep learning components, such as convolutional layers or graph neural networks (which are well-suited to the spatial aspect of the task). Such a model would also be considerably more expensive to train; nonetheless, we expect that most of the conclusions derived for the one-dimensional demonstrator presented in this study will translate into the three-dimensional case.

\section{Methodology}

\subsection{Data}
To train the emulator we used a one-dimensional configuration for the physical General Ocean Turbulence Model \citep[GOTM,][]{burchard2006description} coupled through the Framework for Aquatic Biogeochemical Models \citep[FABM,][] {bruggeman2014general} to the biogeochemical European Regional Seas Ecosystem Model \citep[ERSEM,][]{baretta1995european, butenschon2016ersem}. 

\subsubsection{The physical model GOTM}

GOTM \citep{burchard2006description} is an open-source, one-dimensional water column model designed to simulate the vertical structure and mixing processes of oceans, seas, lakes, and estuaries. It focuses on the effects of turbulence on the transport of momentum, heat, and salinity, providing a flexible framework for testing and comparing different turbulence closure schemes. Because of its modular design and relatively low computational cost, GOTM is widely used in oceanographic research, model development, and coupled Earth system applications to investigate impacts of stratification, mixing, and air–sea interactions \citep[e.g.,][]{torres2006sequential, mattern2010sequential, gharamti2017online, torres2020sensitivity, powley2020sensitivity, ciavatta2025control, higgs2025hybrid}.

\subsubsection{The biogeochemical model ERSEM}

ERSEM \citep{baretta1995european, butenschon2016ersem} is a complex marine biogeochemistry model used for operational forecasts in the North-West European Shelf seas \citep[e.g.,][]{fennel2019advancing}. It describes cycling of multiple elements (e.g. carbon, nitrogen, phosphorus), focusing on lower trophic level dynamics. ERSEM represents four functional types of phytoplankton (diatoms, microphytoplankton, nanophytoplankton and picophytoplankton), three functional types of zooplankton (heterotrophic nanoflaggelates, microzooplankton and mesozooplankton), heterotrophic bacteria, three size-classes of detrital matter and three types of dissolved organic matter (labile, semi-labile and semi-refractory). ERSEM uses variable stoichiometry, with biomass being represented in different currencies, i.e. carbon, nitrogen, phosphorus, in some cases silicon, and chlorophyll-$a$ for phytoplankton. ERSEM includes four inorganic nutrients (nitrate, phosphate, silicate and ammonium),  dissolved oxygen concentrations, includes a carbonate system \citep{artioli2012carbonate}, and a benthic ocean component. The ERSEM configuration used here consists of 52 prognostic pelagic state variables.

\subsubsection{The L4 configuration}

The coupled GOTM-FABM-ERSEM model was configured for the L4 station in the western English Channel. The L4 station is part of the Western Channel Observatory\footnote{\url{https://www.westernchannelobservatory.org.uk/}} at a 50m deep location approximately 13km from the Plymouth Sound. The L4 location (50.25$^{\circ}$N, 4.217$^{\circ}$W) is highly biologically productive with seasonally stratified dynamics \citep{pingree1978tidal}. The station provides long-term  time-series which include nutrients, chlorophyll-$a$ and dissolved oxygen \citep[since 1988,][]{harris2010l4}. The GOTM-FABM-ERSEM L4 configuration, described in \citet{powley2020sensitivity}, uses 100 vertical grid-cells with variable spacing of the grid. The atmospheric forcing for the one-dimensional model was obtained with hourly resolution from the ECMWF ERA-5 reanalysis\footnote{\url{https://www.ecmwf.int/en/forecasts/datasets/reanalysis-datasets/era5}} interpolated into the L4 location. Furthermore, the model uses monthly relaxation in temperature and salinity towards the L4 observations and relaxes the four ERSEM nutrients (nitrate, phosphate, silicate and ammonium) towards seasonal climatology from L4 observations. The latter is intended to prevent spurious long-term trends arising from the absence of atmospheric and riverine nutrient inputs, which are not represented in the one-dimensional model.

\subsubsection{The GOTM-FABM-ERSEM free simulation}

\begin{figure}[h!]
    \centering
    \includegraphics[width=\textwidth]{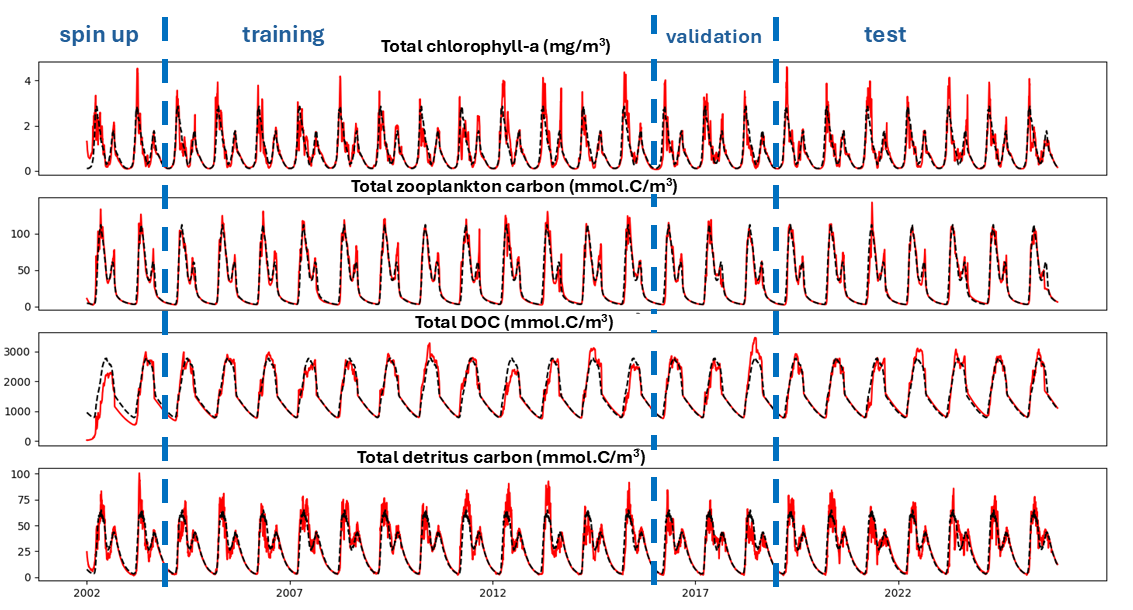}
    \caption{The ocean surface time series for four selected biogeochemistry variables from the 1D GOTM-FABM-ERSEM simulation spanning the 2002-2025 period. The different data (training/validation/test) are clearly marked by the vertical lines.}
    \label{fig:1}
\end{figure}

\begin{figure}[h!]
    \hspace{-2cm}
    \includegraphics[width=18cm]{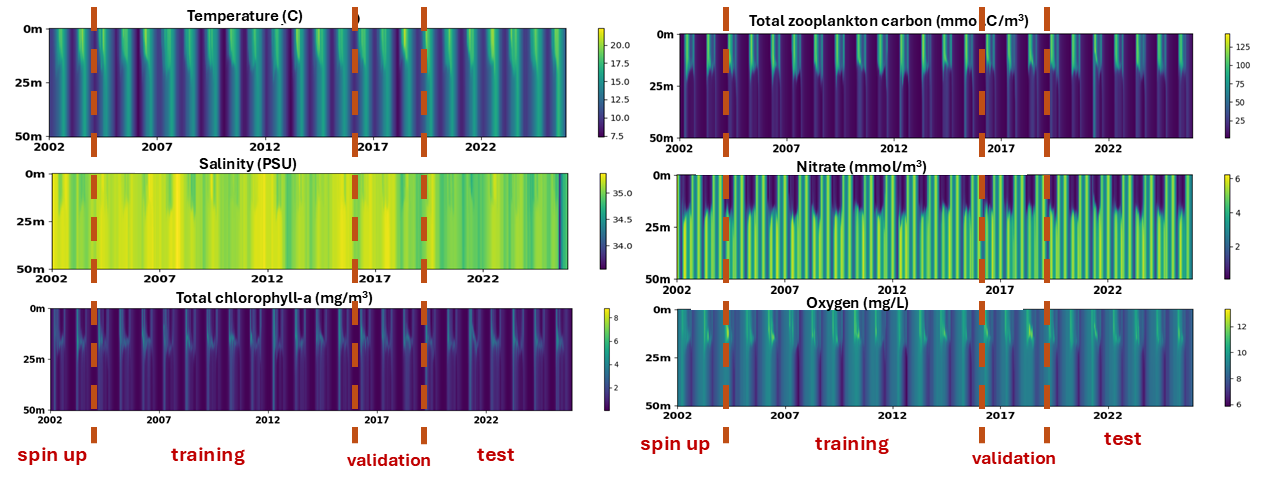}
    \caption{Six selected ocean biogeochemistry and physics variables from the GOTM-FABM-ERSEM simulation, shown in Hovm\"oller plots. As in Fig.\ref{fig:1}, the training, validation and test periods are clearly marked.}
    \label{fig:2}
\end{figure}

The free run data used to train, validate and test the model were obtained from a (nearly) 24 year simulation initialized on 1\textsuperscript{st} February 2002 and running until 31\textsuperscript{st} December 2025. The initial conditions were representative of the ocean state at L4 at the start of the simulation. To minimise any influence of the initialisation, however, the period up to 1\textsuperscript{st} January 2004 (approximately two years) was treated as model spin-up and excluded from the analysis. 
As shown by Fig.\ref{fig:1}-Fig.\ref{fig:2}, this was sufficient. 
As marked in Fig.\ref{fig:1}-Fig.\ref{fig:2} (see also Fig.\ref{fig:A1} in the Appendix), the 12 year period between 1\textsuperscript{st} January 2004 and 31\textsuperscript{st} December 2015 was used to train the emulators, the period between 1\textsuperscript{st} January 2016 and 31\textsuperscript{st} December 2018 was used as validation data, and the 7 years between the 1\textsuperscript{st} January 2019 and 31\textsuperscript{st} December 2025 as the test data. All the data supplied to the deep learning models had a daily resolution.

As shown in Fig.\ref{fig:1}–Fig.\ref{fig:2} and Fig.\ref{fig:A1} in the Appendix, the one-dimensional simulation is characterised by strong seasonal cycle. While this reflects the pronounced seasonal cycle observed at L4, the modelled seasonality is likely exaggerated due to the absence of horizontal transport and riverine inputs. The seasonal dynamics are driven by the winter storms keeping the water column fully mixed. In the Spring, stratification breaks the water column into distinct two layers, and combined with the increased light, triggers the phytoplankton Spring bloom. The Spring bloom consumes the surface nutrients and the surface phytoplankton growth becomes limited by insufficient nutrient concentrations during the Summer, with deep chlorophyll maxima (DCMs) forming around the picnocline (Fig.\ref{fig:2}). The water column gets subsequently mixed by the storms in the late Summer - early Autumn, creating a secondary bloom, after which the growth becomes again light-limited in the Winter. Fig.\ref{fig:1}-Fig.\ref{fig:2} and the Fig.\ref{fig:A1} of the Appendix also show a reasonable level of inter- and intra-annual variability, which is larger for phytoplankton, zooplankton and oxygen variables than for the nutrients (Fig.\ref{fig:1}-Fig.\ref{fig:2}, Fig.\ref{fig:A1} of the Appendix). The small inter-annual variability seen in Winter nutrients can be explained by the fact that nutrients are relaxed towards their seasonal climatology, so when the water column is fully mixed in the Winter, the surface nutrients are roughly the same every year. However inter- and intra-annual variability in nutrients is visible in the Summer, when the surface nutrients are exhausted, as the irregular Summer storms occasionally mix the water column and bring some nutrients into the surface (Fig.\ref{fig:A1} of the Appendix). The strong climatological signal in the time-series of the ERSEM variables represents an easily learnable signal, and this needs to be considered when we evaluate the deep learning model performance.

%Describe in detail the data used, this includes description of L4 and the 1D modelling set-up. Show 2 Figures (Fig.\ref{fig:1} and Fig.\ref{fig:2}), one with the time series and one with Hovmollers. Another Figures in the Appendix (Fig.\ref{fig:A1}). Describe the training-validation-test data, the features used, the characteristics of those data - strong climatology and the need to account for this in the metrics etc.

\subsubsection{The L4 reanalysis}

Apart from the model free run, we have produced a L4 reanalysis, by assimilating a satellite product for total surface chlorophyll-$a$ from Ocean Color Climate Change Initiative \citep[v6.0 of OC-CCI,][]{sathyendranath2019ocean} into the model configuration used in the free run. The available daily OC-CCI satellite chlorophyll data were collected in an area of 100km radius around the L4 location and averaged for each day. This ensured good data availability, such that assimilation of satellite chlorophyll-$a$ concentrations happened throughout the Spring-Summer period nearly every day, with slightly bigger gaps in the Autumn-Winter period. The assimilation was done for the 1\textsuperscript{st} January 2004 - 31\textsuperscript{st} December 2025 period, after the initial 2002-2003 spin up of the GOTM-FABM-ERSEM model. The assimilation was performed with the Ensemble and Assimilation Tool \citep[EAT][]{bruggeman2023eat}, running a 100 member ensemble and using an Ensemble Kalman Filter. The ensemble members were generated by perturbing 15 selected ERSEM parameters based on the sensitivity analysis of \citet{ciavatta2025control}, drawing the parameter values from the same uniform prior probability density functions (PDFs) as in \citet{ciavatta2025control}. Generating the ensemble solely by the ERSEM parameter perturbations ensured that only the ERSEM variables were perturbed, and that the physics remained the same among all the ensemble members. Similarly to the ``RUS'' reference scheme from \citet{higgs2025hybrid}, the ensemble was used to only estimate the background variances and spatial vertical correlations. To mimic the reanalysis generated by the Met Office for the North-West European Shelf, we have assimilated total chlorophyll-$a$ univariately, and partitioned the total chlorophyll-$a$ increments to the ERSEM phytoplankton functional types biomass components using the forecast community structure and stoichiometry \citep[see][]{skakala2018assimilation, higgs2025hybrid}. The observational error was set to 15\% to ensure that the assimilation will have significant impact on the reanalysis state.

The reanalysis was then compiled by taking the 100-member ensemble mean with daily temporal resolution and used to train emulators in an exact analogy to the free run, except for reducing the training and validation data period from 15 to 14 years (maintaining the first 80\% of the time-series as the training and the last 20\% as the validation data). Starting the training and validation data period one year later (on the 1\textsuperscript{st} January 2005 instead of 1\textsuperscript{st} January 2004) allowed us to leave an extra year to spin-up the assimilation run before the reanalysis data were used to train the deep learning models.

\subsection{The deep learning architectures}

\subsubsection{The LSTM emulator}

A Long Short-Term Memory (LSTM) neural network \citep{hochreiter1997long} was developed to emulate the temporal dynamics of selected surface and vertically integrated state variables from the GOTM-FABM-ERSEM one-dimensional water column biogeochemical model. These variables were selected based on both, their importance for the model stakeholders and their relevance for the ecosystem dynamics, i.e. the reduced model is to a degree a refined version of the low-complexity, nutrients-phytoplankton-zooplankton (NPZ) models. The emulator predicts the state of the system at time $t$ using a history of previous biogeochemical states and environmental forcing variables. The emulated time-steps are daily, as opposed to the 15 minute time step of the GOTM-FABM-ERSEM simulator, or the 5 minute time step of the three-dimensional UK regional model.

The LSTM is using as inputs forcing variables, split into a) atmospheric forcing variables: photosynthetically active radiation (PAR), meridional and zonal wind components, precipitation and surface heat flux, and b) ocean physics forcing variables: sea surface temperature (SST), sea surface salinity (SSS), mixed layer depth (MLD). Furthermore, it uses c) a range of biogeochemistry variables: surface chlorophyll-$a$ concentration for four ERSEM phytoplankton functional types (PFTs), four surface nutrient concentrations (nitrate, phosphate, silicate, ammonium), three zooplankton functional type surface carbon concentrations, dissolved surface oxygen concentration, dissolved sea bottom oxygen concentration (averaged across the bottom 5 meters), vertically averaged concentrations of the four PFTs chlorophyll-$a$, as well as the three zooplankton functional types carbon. The atmospheric forcing variables were taken from the daily averages of the ERA-5 product that forced the GOTM-FABM-ERSEM simulator and the oceanic forcing inputs were taken from the daily-resolution GOTM-FABM-ERSEM simulation outputs. For each prediction, the model receives a sequence of the previous 30 daily average values (lookback window = 30) of both state and forcing variables. In addition, the forcing variables at the prediction time step are supplied separately, reflecting the assumption that forcing at the predicted day is known. The LSTM then predicts the selected biogeochemistry variables (those listed in point c) at the next daily time step. As the primary role of the emulators is to predict anomalies, we also tested an LSTM architecture in which the emulator used daily anomaly data, with the day of the year included as an additional input. However, this design did not perform as well as the one described above. A possible explanation is that the high-dimensional function the emulator is attempting to learn is more naturally represented by the total values rather than by the anomalies.

Training samples were created using a sliding window approach. The inputs and outputs were normalized using z-score normalization ($\mu$ is mean and $\sigma$ the standard deviation):
\[x\rightarrow \frac{x-\mu}{\sigma}.\]
To avoid information leakage, normalization statistics (mean and standard deviation) were computed using only the training subset and subsequently applied to both training and validation datasets. State variables, forcing variables, and prediction targets were normalized independently.

The LSTM model architecture is described in Table~\ref{tab:lstm_architecture}; the emulator is structured as a multi-input neural network architecture. Combined input sequences are processed by a single LSTM layer containing 64 hidden units. The LSTM acts as a temporal encoder, extracting information from the previous month of system evolution and environmental forcing. Only the final hidden state is retained, providing a compact representation of the recent system history. The resulting latent representation is concatenated with the forcing vector at the prediction time step. This combined feature vector is passed through: a) a fully connected layer with 64 neurons and ReLU activation, b) a linear output layer with 20 neurons corresponding to the predicted biogeochemical state variables.

\begin{table}[htbp]
\centering
\begin{tabular}{lllll}
\textbf{Step} & \textbf{Layer Type} & \textbf{Input Shape} & \textbf{Output Shape} & \textbf{Details} \\
\hline \hline
1 & Input (bio) & $(\text{lookback},\, n_{\text{state}})$ & same & Past state variables \\
2 & Input (forcing) & $(\text{lookback},\, n_{\text{forcing}})$ & same & Past forcing variables \\
3 & Input (present) & $(n_{\text{forcing}},)$ & same & Present forcing (single step) \\
4 & Concatenate & $(\text{lookback},\, n_{\text{state}} + n_{\text{forcing}})$ & same & Combine past sequences \\
5 & LSTM (64) & $(\text{lookback},\, \text{features})$ & $(64,)$ & Sequence encoding \\
6 & Concatenate & $(64 + n_{\text{forcing}})$ & same & Add present forcing \\
7 & Dense (ReLU) & $(64 + n_{\text{forcing}})$ & $(64,)$ & Fully connected layer \\
8 & Dense (output) & $(64,)$ & $(n_{\text{state}},)$ & Final prediction \\
\end{tabular}
\vspace{.3cm}
\caption{Architecture of the LSTM-based forecasting model.}
\label{tab:lstm_architecture}
\end{table}

The model was trained using the Adam optimizer and a mean absolute error (MAE) loss function. Training was performed with a batch size of 512 samples and without shuffling to preserve the temporal ordering of the data. Two callback mechanisms were used: a) early stopping with a patience of 20 epochs to prevent overfitting, b) learning rate reduction on plateau, halving the learning rate after five epochs without validation loss improvement, with a minimum learning rate of $10^{-4}$. Each model was trained for a maximum of 250 epochs.
To assess the variability arising from random initialization, an ensemble of 15 independently trained LSTM models was generated. The weights and training histories of all ensemble members were saved for subsequent analysis. Fig.\ref{fig:A1.1} in the Appendix shows the training curves for the LSTM ensemble.

%Describe design, 15 member ensemble, cost function, normalization, initialization, lookback window... {\bf Show training performance (loss) in the Appendix? Maybe as 2 panels, 1 for LSTM and other as CNN showing the ensemble median (with quartiles) loss function curves?}

\subsubsection{The physics-informed 1D CNN emulator}

A one-dimensional physics-informed convolutional neural network (1D CNN, \cite{lecun1995cnn}) based on a ResNet architecture was also developed to emulate the evolution of the full vertical state of the GOTM-FABM-ERSEM water column model. Unlike the reduced-order LSTM emulator, which predicts selected surface and vertically averaged variables, the CNN predicts the complete vertical profiles of all simulated pelagic state variables. This emulator does not reduce the number of degrees of freedom of the simulator, but saves computational power by reducing the simulator time-step to a daily resolution.

The physics-informed 1D CNN emulator predicts daily averages of 52 ERSEM pelagic variables across the 100 vertical layers of the GOTM-FABM-ERSEM model outputs. It takes the same daily averaged atmospheric forcing variables as the LSTM network (originating from ERA-5) and daily averages of temperature and salinity across the full water column from the GOTM-FABM-ERSEM simulation. The information about the depth of the vertical layers is not provided directly to the emulator, i.e. when this information was included, the performance of the emulator became worse, possibly due to overfitting.

The CNN is designed to learn local vertical interactions while preserving spatial structure throughout the water column, as described in Table~\ref{tab:residual_model_architecture}. The forcing vector is first broadcast across all vertical layers using a repeat operation. This forcing field is concatenated with the vertical state profiles to form the network input. An initial one-dimensional convolutional layer with 128 filters and kernel size 5 projects the input into a higher-dimensional latent space. The core of the architecture consists of six residual convolutional blocks. Each block contains: (a) a one-dimensional convolution with 128 filters and kernel size 5, (b) a Gaussian Error Linear Unit (GELU) activation, (c) a second one-dimensional convolution with 128 filters, (d) a residual skip connection that adds the block input to its output. The residual formulation enables efficient training of deeper networks while maintaining gradient flow and preserving fine-scale vertical information.

\begin{table}[htbp]
\centering
\resizebox{\textwidth}{!}{
\begin{tabular}{lllll}

\textbf{Step} & \textbf{Layer Type} & \textbf{Input Shape} & \textbf{Output Shape} & \textbf{Details} \\
\hline\hline
1 & Input (state) & $(d,\, n_{\text{vars}})$ & same & Vertical profile of state variables \\
2 & Input (forcing) & $(n_{\text{forcing}},)$ & same & External forcing vector \\
3 & RepeatVector & $(n_{\text{forcing}},)$ & $(d,\, n_{\text{forcing}})$ & Broadcast forcing over depth \\
4 & Concatenate & $(d,\, n_{\text{vars}} + n_{\text{forcing}})$ & same & Combine state and forcing \\
5 & Conv1D (128) & $(d,\, \text{features})$ & $(d,\, 128)$ & Initial projection (kernel size = 5) \\
6 & Residual Block $\times 6$ & $(d,\, 128)$ & $(d,\, 128)$ & Two Conv1D + GELU with skip connection \\
7 & Conv1D (output) & $(d,\, 128)$ & $(d,\, n_{\text{vars}})$ & Predict state tendency ($dx$) \\
8 & Add (residual) & $(d,\, n_{\text{vars}})$ & $(d,\, n_{\text{vars}})$ & $x + dx$ (state update) \\

\end{tabular}}
\caption{Architecture of the residual 1D CNN model with depth-wise processing.}
\label{tab:residual_model_architecture}
\end{table}

The network predicts a state tendency field
$\Delta \mathbf{X}_t$, using a final (1$\times$1) convolutional layer that maps the latent representation back to the original variable space. Rather than directly predicting the next state, the model uses a residual update: $\mathbf{X}_t + \Delta \mathbf{X}_t.$
This formulation allows the network to learn temporal increments rather than absolute states, which generally improves numerical stability and training efficiency. 

To improve long-term forecasting skill, the emulator was trained autoregressively using multi-step rollouts. Training sequences of 10 days were generated from the model output. For each sequence, the emulator receives the initial state ($\mathbf{X}_t$) and the corresponding forcing sequence $\mathbf{F}_t,\mathbf{F}_{t+1},\ldots,\mathbf{F}_{t+9}$. The emulator predicts the next state, which is then recursively fed back into the model to generate subsequent predictions. The loss is accumulated across the entire rollout horizon:
\[
\mathbf{X}_{t}
\rightarrow
\hat{\mathbf{X}}_{t+1}
\rightarrow
\hat{\mathbf{X}}_{t+2}
\rightarrow \cdots \rightarrow
\hat{\mathbf{X}}_{t+10}.
\]
This training strategy encourages the emulator to remain stable and accurate when used iteratively in forecasting mode.

Similarly to the LSTM, the 1D CNN inputs and outputs were normalized using z-score normalization. The objective loss function combines prediction accuracy with physically motivated constraints. The primary loss is the mean absolute error between predicted and simulated states, but additional penalties are imposed to discourage spurious changes in the biogeochemistry state. These are imposed through two types of physics constrains: (i) mass conservation in carbon, nitrogen and phosphorus, and (ii) bounding the interval of values for each variable from below by the zero (after the z-score normalization this equals to negative of the variable's mean divided by its standard deviation) and from above by 20 times the variable's mean. The total MAE loss function $\mathcal{L}$ was defined as:
\[\mathcal{L} = \mathcal{L}_{pred} + \frac{\mathcal{L}_{mass} + \mathcal{L}_{bound}}{3}\]
with
\begin{equation}\label{pred}
\mathcal{L}_{pred} = \langle |\hat{\mathbf{X}}_{pred} - \hat{\mathbf{X}}_{train}|\rangle 
\end{equation},
\begin{equation}\label{mass}
\mathcal{L}_{mass} = \frac{|M^{C}_{t} - M^{C}_{t-1}|}{M^{C}_{t-1}} + \frac{|M^{N}_{t} - M^{N}_{t-1}|}{M^{N}_{t-1}} + \frac{|M^{P}_{t} - M^{P}_{t-1}|}{M^{P}_{t-1}},  
\end{equation}
\begin{equation}\label{bound}
\mathcal{L}_{bound} = \left\langle\left\langle max\left(\hat{\mathbf{X}}_{pred} - \hat{\mathbf{X}}_{up}, 0\right)^{2} + min\left(\hat{\mathbf{X}}_{pred} - \hat{\mathbf{X}}_{low}, 0\right)^{2}\right\rangle_{vert}\right\rangle_{var}. 
\end{equation}
In the Eq.\ref{mass}, $M^{C}_{t}, M^{N}_{t}, M^{P}_{t}$ are the total integrated carbon ($M^{C}_{t}$), nitrogen ($M^{N}_{t}$) and phosphorus ($M^{P}_{t}$) masses at the day $t$. In the Eq.\ref{bound}, $\hat{\mathbf{X}}_{up}$ and $\hat{\mathbf{X}}_{low}$ are upper and lower bounds of the normalized variable $\hat{\mathbf{X}}$, applied as constants across the vertical water column. The ``$\rangle_{vert}$'' and ``$\rangle_{var}$'' symbols denote averaging across vertical dimension and the variables. The physical constraints were imposed through the loss function during training, and the same bounds on the values used during training were also enforced during prediction.

Training was performed using the Adam optimizer with an initial learning rate of $10^{-3}$ and gradient clipping to improve numerical stability during multi-step rollout training. Mini-batches of 32 samples were used, and training proceeded for up to 50 epochs. Validation loss was evaluated after each epoch, and early stopping was applied when no improvement was observed for ten consecutive epochs. The best-performing model weights were saved according to the validation loss, with the training and validation curves shown in Fig.\ref{fig:A1.2} of the Appendix. 
%Unlike LSTM, different initialization values led to significant differences in network performance, i.e. when an ensemble of models was trained differing in the randomly allocated parameter values during the training process, some ensemble members showed stable behavior and high quality forecasts, whilst others demonstrated unstable behavior and low skill. For 1D CNN we therefore present a single well performing model, but keep in mind that future work should be dedicated to make the training more robust.
Unlike the LSTM, the 1D CNN was highly sensitive to the stochastic elements of the training procedure. Different random weight initializations and training/validation batch splits led to substantial differences in model performance. Specifically, when an ensemble of models was trained using different random seeds, some ensemble members exhibited stable behavior and produced high-quality forecasts, whereas others showed unstable behavior and low skill. We therefore present results for a single well-performing 1D CNN model. However, future work should focus on improving the robustness of the training process.
%Unlike the LSTM, the 1D CNN was highly sensitive to weight initialization. Different initialization values led to substantial differences in model performance; that is, when an ensemble of models was trained using different random hyperparameter choices, some ensemble members exhibited stable behavior and produced high-quality forecasts, whereas others showed unstable behavior and low skill. We therefore present results for a single well-performing 1D CNN model. However, future work should focus on improving the robustness of the training process.
%As in case of LSTM, to quantify uncertainty associated with neural network initialization, an ensemble of 15 independently trained CNN emulators was generated using different random seeds.

%Similarly, the design, 10 day rollout training, physics constrains, 15 member ensemble ... 

\subsection{The experiments}

We have performed a number of experiments addressing both long-term decadal and short-term prediction with both types of emulators:
\begin{itemize}
\item {\bf Short-range free-run prediction} experiments where emulators, trained on the free run, were initialized at each day across several years of the test data period and for each day's initialization a 10-day emulator forecast was run.
\item {\bf Decadal free-run prediction} experiments, where both types of emulators, trained on the free run, were initialized on the 1\textsuperscript{st} January 2004 and run for 22 years until the 31\textsuperscript{st} December 2025. These emulator runs span through all training, validation and test data period.
\item {\bf Seven-year reanalysis forecast} experiments where both emulators were trained on the L4 reanalysis data for the 1\textsuperscript{st} January 2005 - 31\textsuperscript{st} December 2018 period and then initialized from the reanalysis on the 1\textsuperscript{st} January 2019 and compared with a GOTM-FABM-ERSEM simulator free run initialized on the same day from the same reanalysis. Both emulators and the free run were run for the 1\textsuperscript{st} January 2019 - 31\textsuperscript{st} December 2025 period.
\item {\bf Decadal free run forecasting with noise} experiments were performed, where time-autocorrelated multiplicative red noise (correlated roughly on a 15 day time-scale) was independently applied to the model inputs and impact of the noise on emulator simulation was investigated. For CNN the noise was applied as constant across the water column. These experiments were run again for the 1\textsuperscript{st} January 2004 to the 31\textsuperscript{st} December 2025 period.  
\end{itemize}

%Describe decadal prediction and short-range prediction experiments. Describe reanalysis-based experiment, data used there, the free run initialized from the reanalysis etc. Show in a Table indicators selected in this paper (i.e. the predicted variables analysed in the paper and their definitions).

\subsection{Validation metrics}
In evaluating the emulators, we focus on a range of outputs listed in Tab.\ref{tab:variables}. These variables are of particular interest for higher-trophic-level dynamics, the carbon cycle and carbon export, hypoxia, and biological productivity.

\begin{table}[htbp] \centering \caption{Summary of the variables used to validate the emulators. The columns two and three list which variables are used for which emulators. The abbreviations that need explaining are: dissolved organic carbon - DOC, dissolved inorganic carbon - DIC.} \label{tab:variables} \begin{tabular}{|c|c|c|}
\hline
{\bf Variable} & {\bf LSTM} & {\bf 1D CNN} \\ \hline
Surface total chlorophyll-$a$ & yes & yes \\
\hline
Surface total zooplankton carbon & yes & yes \\
\hline
Surface nitrate & yes & yes \\
\hline
Surface phosphate & yes & yes \\
\hline
Surface silicate & yes & yes \\
\hline
Surface oxygen & yes & yes \\
\hline
Vertically averaged total chlorophyll-$a$ & yes & yes \\
\hline
Vertically averaged total zooplankton carbon & yes & yes
\\
\hline
Vertically averaged total detritus carbon & no & yes \\
\hline
Vertically averaged total DOC & no & yes \\
\hline
Vertically averaged total DIC & no & yes \\
\hline
Sea bottom oxygen & yes & yes \\
\hline
\end{tabular}
\end{table}

Due to the strong seasonal signal in the data (e.g. Fig.\ref{fig:1}-Fig.\ref{fig:2}) we need to exclude the possibility that the deep learning emulators converged to a solution where they learned purely the seasonal climatology. Such solution could score well in terms of the loss function, but is meaningless from the point of emulating the process-based simulator dynamics. To exclude such possibility, we derived for the variables from Tab.\ref{tab:variables} daily anomalies, by calculating for each variable daily climatology across the 15-year, 2004-2018, period and subtracting the climatology from its time-series. 

For validation by the test data we used established metrics, such as the mean absolute error (MAE) and root mean squared error (RMSE), which gave generally very similar results. These metrics, when used to compare the anomalies in emulated ($y_{emu}$) and test ($y_{test}$) data, were normalized by the size of test data anomalies as
\begin{equation}\label{eq:RMSE}
\hbox{RMSE}_{norm} = \frac{\sqrt{\langle (y_{emu} -y_{test})^{2}\rangle}}{\sqrt{\langle y_{test}^{2}\rangle}}
\end{equation}
and   
\begin{equation}\label{eq:MAE}
\hbox{MAE}_{norm} = \frac{\langle |y_{emu} -y_{test}|\rangle}{\langle |y_{test}|\rangle}.
\end{equation}

Similarly, when we evaluated short-range forecasts, the RMSE (or MAE) forecast skill was normalized by the RMSE (or MAE) skill of persistence, which is prediction based on the assumption that the current state will persist into the future. This provides a measure for the model skill relative to the natural rate of dynamics of the system. 

The RMSE and MAE metrics produced qualitatively similar results. In this study, we present MAE when evaluating predictions of total values across longer time-scales and RMSE when evaluating predictions of anomalies, or short-medium (10-day) range predictions. RMSE places greater emphasis on larger errors and therefore highlights model performance during periods with high-magnitude values, whereas MAE provides a more balanced assessment across the full time series. This sensitivity to large deviations is particularly relevant for anomalies, as it emphasizes the extreme events that are often of greatest predictive interest. Similarly, for short- to medium-range forecasts, model performance during highly dynamic conditions (weighted higher by RMSE) is generally of greater interest than performance during relatively quiescent periods, when simple persistence can already provide accurate predictions. Consequently, although the emulators were trained using an MAE loss function to promote balanced performance across all conditions, anomaly predictions and short-medium range predictions are evaluated using RMSE.

\subsection{Emulator explainability}
To assess the influence of individual input variables on the predictions, saliency analysis is applied to the trained LSTM model. This method quantifies the sensitivity of each predicted output to perturbations in the input features by computing gradients through the network. Specifically, the partial derivative of a predicted output with respect to each element of the input vector is evaluated using backpropagation. For a predicted output $\hat{y}$ and an input variable $x_{i,t-\tau}$, where $i$ denotes the input index and $\tau$ denotes the lag in days prior to the forecast step, the sensitivity is defined as:

\begin{equation}
    S_{i,\tau} = \left| \frac{\partial \hat{y}}{\partial x_{i,t-\tau}} \right|
\end{equation}

\noindent Taking the absolute value ensures that both positive and negative sensitivities contribute equally to the magnitude, reflecting the strength of the dependency regardless of its sign. This is computed for each forecast initialisation time $t$ across the full evaluation period, yielding a time-varying sensitivity field for each input-output pair.

For historical state variables, sensitivities are computed across the full 30-day lookback window, producing a sensitivity profile as a function of lag $\tau \in [0, 30]$ days. For meteorological forcing variables prescribed at the forecast step (i.e. $\tau = 0$), sensitivity is computed with respect to the concurrent forcing value provided as input to the model for the forecast step being evaluated.

To summarise the temporal evolution of sensitivities, heatmaps are constructed with the full 
evaluation period time series on the horizontal axis and input variable on the vertical axis, 
where each column represents the peak sensitivity across the 30-day lookback window for that 
forecast date.
We then characterise the effective memory timescale of the model for a given input-output pair, the lag at which the mean sensitivity first drops to a given percentage of its peak mean value is identified, with thresholds set at 50\%, 10\%, and 5\%. This provides a concise summary of how far back in time each input variable meaningfully influences the prediction.

The saliency approach adopted here is closely analogous to the adjoint method widely used in 4D variational data assimilation \citep[e.g.][]{talagrand1987variational, errico1997adjoint}. In 4D variational data assimilation, the adjoint of the forecast model is used to propagate sensitivity information backwards in time from a scalar cost function, typically a measure of forecast error, to the initial conditions or boundary forcings. The resulting adjoint sensitivities quantify how perturbations to the model state at earlier times influence the cost function at the verification time, and are used to compute the gradient required for optimising the initial conditions. The saliency gradients computed here serve an equivalent role: they propagate sensitivity information backwards through the LSTM from a given output, through the sequence of hidden states, to each element of the input window. The analogy is particularly relevant in the context of data assimilation applications, where the sensitivity timescales identified here could inform the specification of observation influence windows or the structure of background error covariance matrices, providing a data-driven estimate of the effective memory of the biogeochemical system as represented by the trained model.

\section{Results and Discussion}

\subsection{Decadal prediction}

\begin{figure}[h!]
    \hspace{-2.5cm}
    \includegraphics[width=18cm]{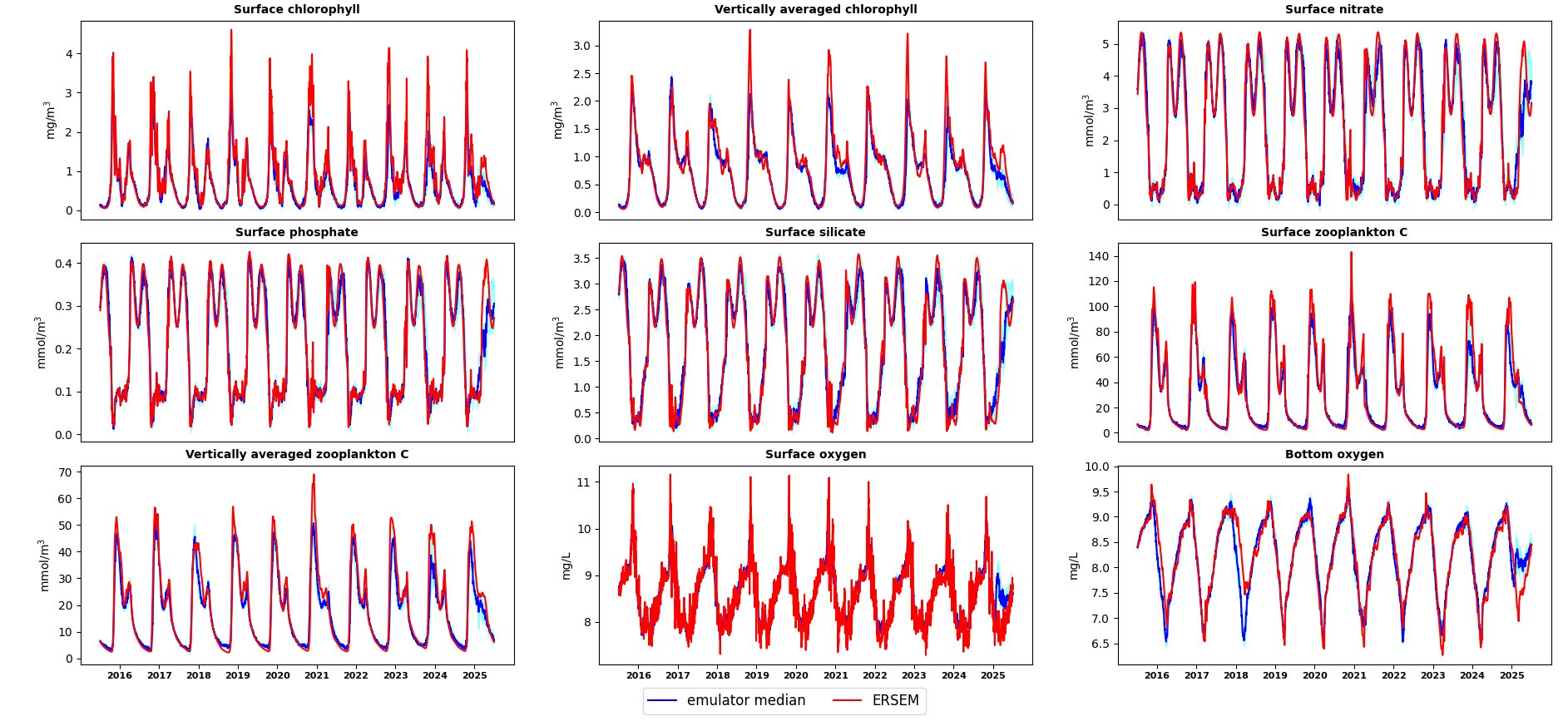}
    \caption{The LSTM prediction of nine selected variables (see Tab.\ref{tab:variables}) across the validation (2016-2018) and test (2019-2025) data period. The predicted LSTM ensemble median value is in blue with two quartiles around the median in aqua color. The simulator validation and test data are in red.}
    \label{fig:3}
\end{figure}

\begin{figure}[h!]
    \hspace{-3cm}
    \includegraphics[width=18cm]{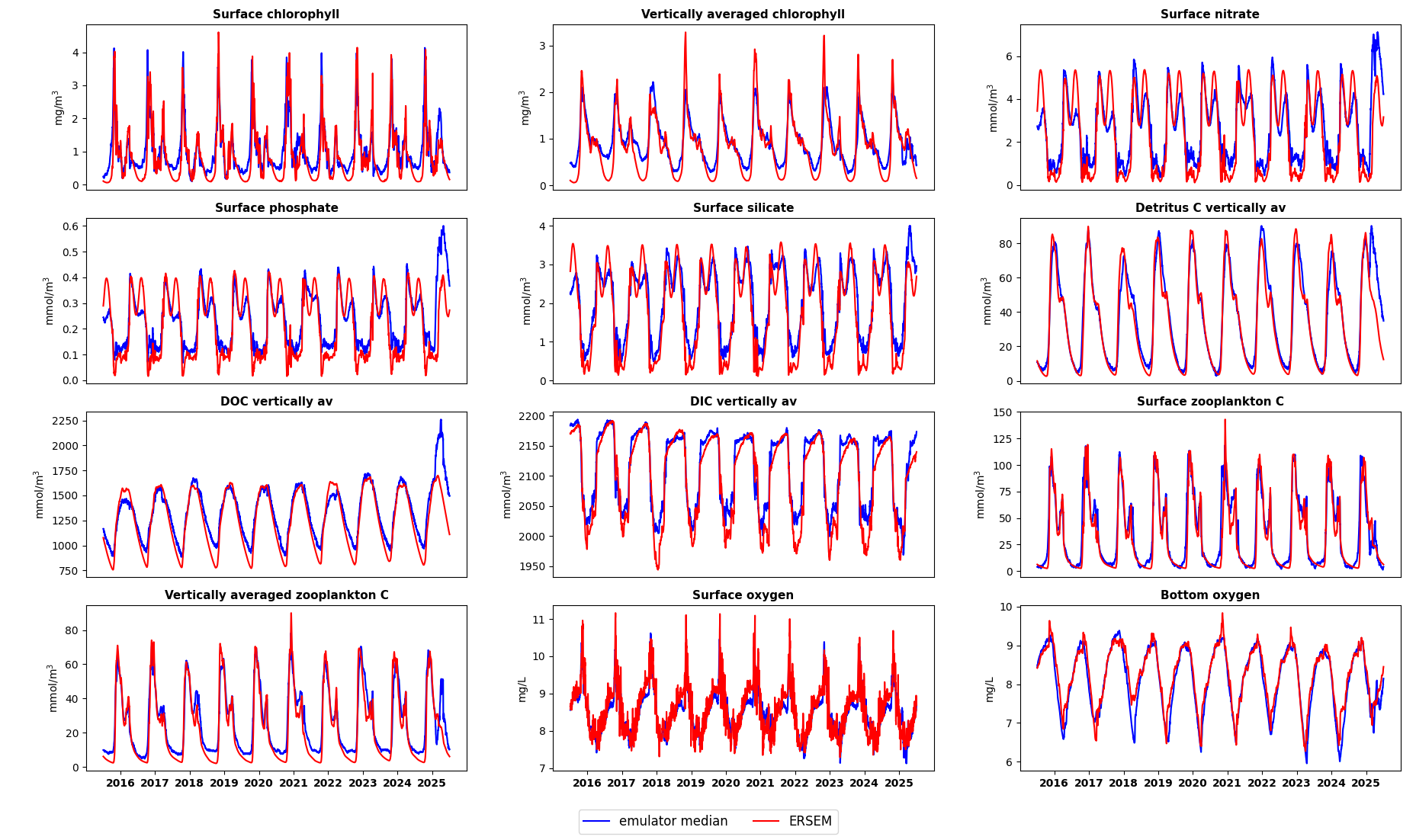}
    \caption{Similar to Fig.\ref{fig:3}, but simulated by the physics-informed 1D CNN emulator. The plotted variables are defined in Tab.\ref{tab:variables}.}
    \label{fig:4}
\end{figure}

\begin{figure}[h!]
    \vspace{-1cm}
    \hspace{-2.5cm}
    \includegraphics[width=18cm, height=18cm]{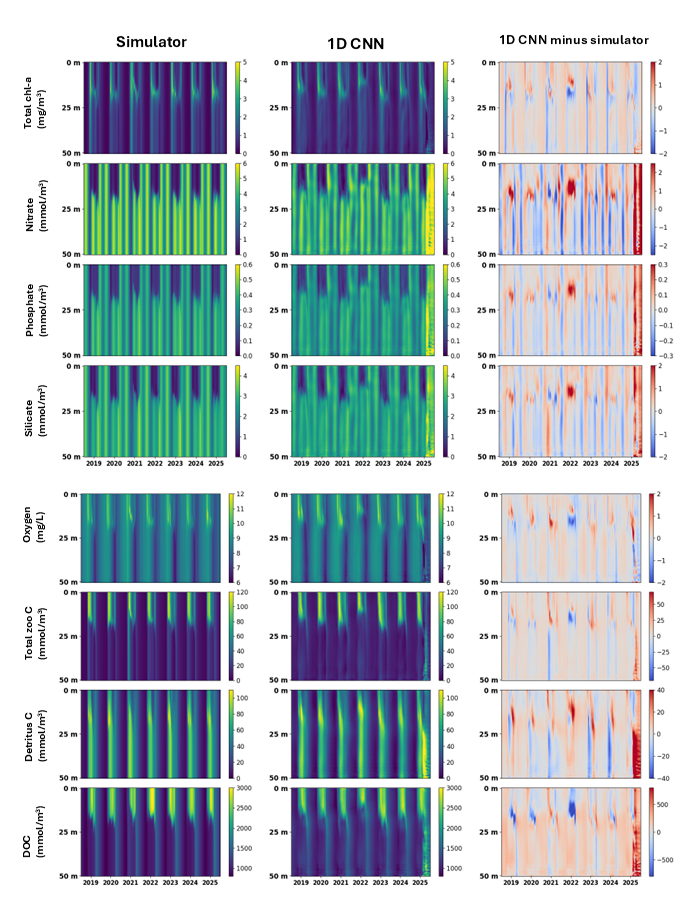}
    \caption{Hovm\"oller plots compared for eight selected variables across the test period (2019-2025) between the test data (left-hand panels) and the 1D CNN emulator (middle column). The difference between those two (emulator minus test data) is shown in the right-hand panels.}
    \label{fig:5}
\end{figure}

We have tested how the emulators, trained on the GOTM-FABM-ERSEM free run simulation, perform on the 22 year forecast (2004-2025) time-scale. The LSTM emulator is stable on these long time-scales and closely reproduces the GOTM-FABM-ERSEM simulator, including during the 2016-2018 validation data and the 2019-2025 test data period (Fig.\ref{fig:3}). The spread of the LSTM ensemble is fairly negligible (Fig.\ref{fig:3}) indicating that the training process converges to a similar solution. The physics-informed 1D CNN simulation is also stable on longer time-scales, until mid-2025  (Fig.\ref{fig:4}-Fig.\ref{fig:5}), when the emulator starts rapidly diverging from the GOTM-FABM-ERSEM simulator trajectory. We assume this behavior is triggered by the salinity anomaly in the GOTM-FABM-ERSEM simulation, which appears around the Summer 2025 (see Fig.\ref{fig:2}). A much more modest effect of this anomaly on the LSTM ensemble can also be observed (Fig.\ref{fig:3}).

The 1D CNN emulator is also able to capture the vertical structure of the GOTM-FABM-ERSEM simulation (Fig.\ref{fig:5}). However, the emulated vertical profiles are notably smoother than those produced by the simulator (Fig.\ref{fig:5}). This behaviour is consistent with the known tendency of autoregressively rolled-out, MAE-trained neural forecasting models to regress towards the conditional mean of plausible future states, resulting in statistically smooth but often dynamically unrealistic trajectories that suppress sharp gradients while preserving lower-frequency, more predictable signals \citep{lam2023learning, bonavita2024some}. This smoothing effect is particularly evident in the representation of the deep chlorophyll maximum (DCM), where the emulator begins to diffuse sharp gradients within the first 10 forecast days (Fig.\ref{fig:A2.5} of the Appendix). Furthermore, there is considerable inter-annual variability in the emulated vertical profiles (Fig.\ref{fig:5}). This spurious variability emerges from error accumulation in the emulator rollout, despite the fact that vertical profiles of ocean physical variables are provided to the emulator each day.

\begin{figure}[h!]
    \vspace{-1cm}
    \hspace{-3cm}
    \includegraphics[width=20cm, height=14cm]{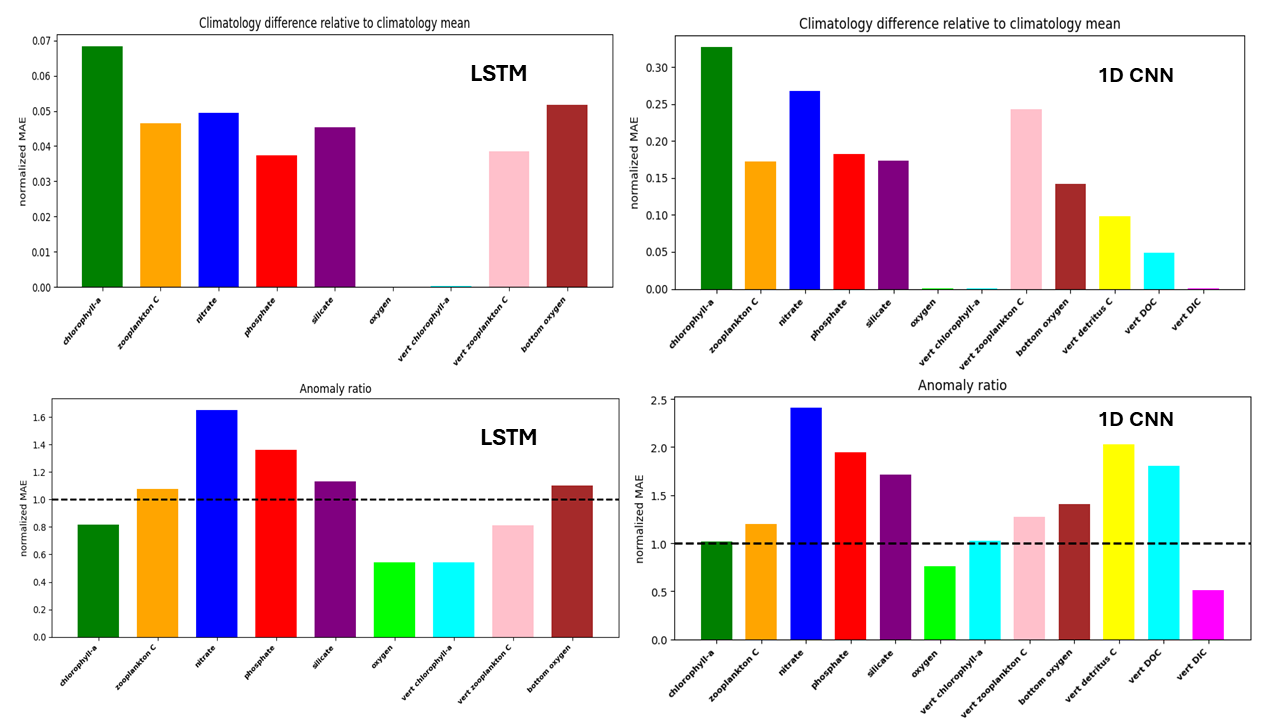}
    \caption{The upper-row bar-plots show the values of mean absolute differences between the emulated seasonal climatology and the seasonal climatology from the training data. The values were normalized by the mean climatology value from the training data and are shown for a range of selected variables. The bottom-row plots show the ratio of mean absolute values of daily anomalies from the emulator runs and from the test data (both across the 2019-2024 period, skipping the year 2025 to avoid the values being impacted by the degraded emulator run in that year). The left-hand panels are the median of the LSTM emulator ensemble and the right-hand panels are the 1D CNN emulator.}
    \label{fig:6}
\end{figure}

The magnitude of the inter- and intra-annual variability simulated by the emulators is broadly comparable to that of the GOTM-FABM-ERSEM simulation (Fig.\ref{fig:6}). This indicates that the emulators have not simply learned the seasonal climatology, but are also capable of producing anomalies, a key requirement for meaningful forecasting. 
Furthermore, Fig.\ref{fig:7} and Fig.\ref{fig:A3} in the Appendix show that both the LSTM and the 1D CNN emulators capture the temporal signatures of these anomalies reasonably well (i.e. for many variables the Pearson correlation, $R$, lies between 0.4 and 0.8). Fig.\ref{fig:7} and Fig.\ref{fig:A3} also show a decreasing trend in emulator skill as a function of forecast lead time (in years), across the validation and test data period. This is expected, as with the increased lead time the emulator moves further away from the data it was trained on.  These results suggest that the deep learning emulators respond to variations in the physical environment in a manner that is largely consistent with the process-based model. Fig.\ref{fig:8} and Fig.\ref{fig:A4} of the Appendix demonstrate how well the size of anomalies can be forecast by the emulators on a decadal scale as a function of lead time (in years). The Figures compare the prediction (RMSE) skill with the scale of the anomalies (see the normalized RMSE in Eq.\ref{eq:RMSE}), which amounts to the RMSE skill of a zero-anomaly prediction model. It is shown that for most of the variables the emulator forecasts outperforms the zero-anomaly prediction during the validation and the first years of the test data period. However the degradation of the model skill with lead time means that in the last years of the test data period (i.e. 2023-2025) the anomaly prediction RMSE skill is worse than the RMSE skill of the zero-anomaly prediction model (Fig.\ref{fig:8} and Fig.\ref{fig:A4}). Further insight into the emulator dynamics can also be obtained from the power spectra (Fig.\ref{fig:AFourier} of the Appendix), which show good agreement between the emulators and the simulator, although some minor smoothing at shorter temporal scales (typically $<$10 days, and occasionally extending to $<$100 days) is apparent in the emulator output.

\begin{figure}[h!]
%    \vspace{-1cm}
    \hspace{-1.5cm}
    \includegraphics[width=16cm]{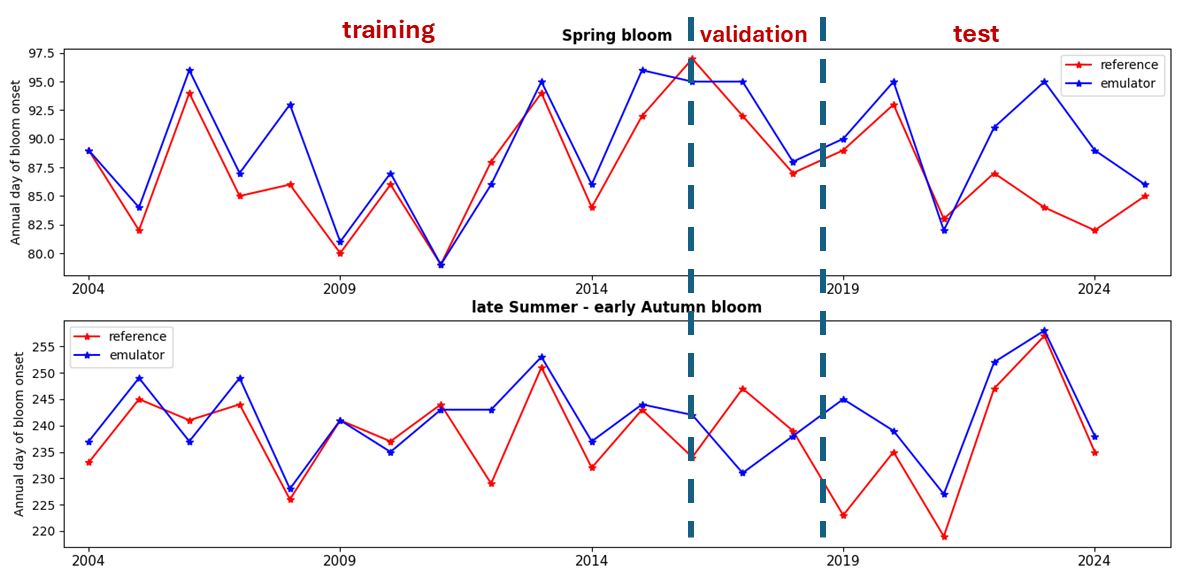}
    \caption{The timing of bloom onset, expressed as the day of year, is shown for the Spring bloom (upper panel) and the second bloom occurring in late Summer to early Autumn (lower panel). The Spring bloom onset was defined as the date when surface total chlorophyll-a first exceeded a threshold of $0.5~ mg.m^{-3}$. The onset of the second bloom was defined as the date when surface total chlorophyll-a exceeded the same threshold above the minimum chlorophyll concentration simulated between the two blooms. To reduce short-term variability, the total surface chlorophyll-a time series used to determine bloom onset was low-pass filtered using a weekly median filter. The GOTM-FABM-ERSEM simulator (red) is compared with the median of the LSTM ensemble (blue). In the bottom plot showing the second bloom we have not shown the 2025 data, since the LSTM simulation was diverging from the ERSEM trajectory and largely missed the second bloom in that last year.}
    \label{fig:6.5}
\end{figure}

Of particular interest is the ability to forecast specific events that have a major impact on marine biogeochemistry in temperate seas, such as phytoplankton blooms. At L4, two blooms typically occur each year: a larger Spring bloom and a smaller late-Summer, early-Autumn bloom (Fig.\ref{fig:1}). Fig.\ref{fig:6.5} shows the skill of the LSTM in forecasting the inter-annual variability in the onset timing of both blooms. The onset of the Spring bloom varies by approximately 2–3 weeks between years, and the LSTM reproduces this variability well throughout the training and validation periods, as well as during part of the test period, with the exception of 2022–2024. The timing of the second bloom exhibits a similar degree of inter-annual variability and is also predicted reasonably well by the LSTM (Fig.\ref{fig:6.5}). These results suggest that emulators driven by ocean physics forecast could be used on a multi-year-to-decadal basis to forecast the onset of algae blooms.

\begin{figure}[h!]
    \vspace{0cm}
    \hspace{-3cm}
    \includegraphics[width=20cm, height=12cm]{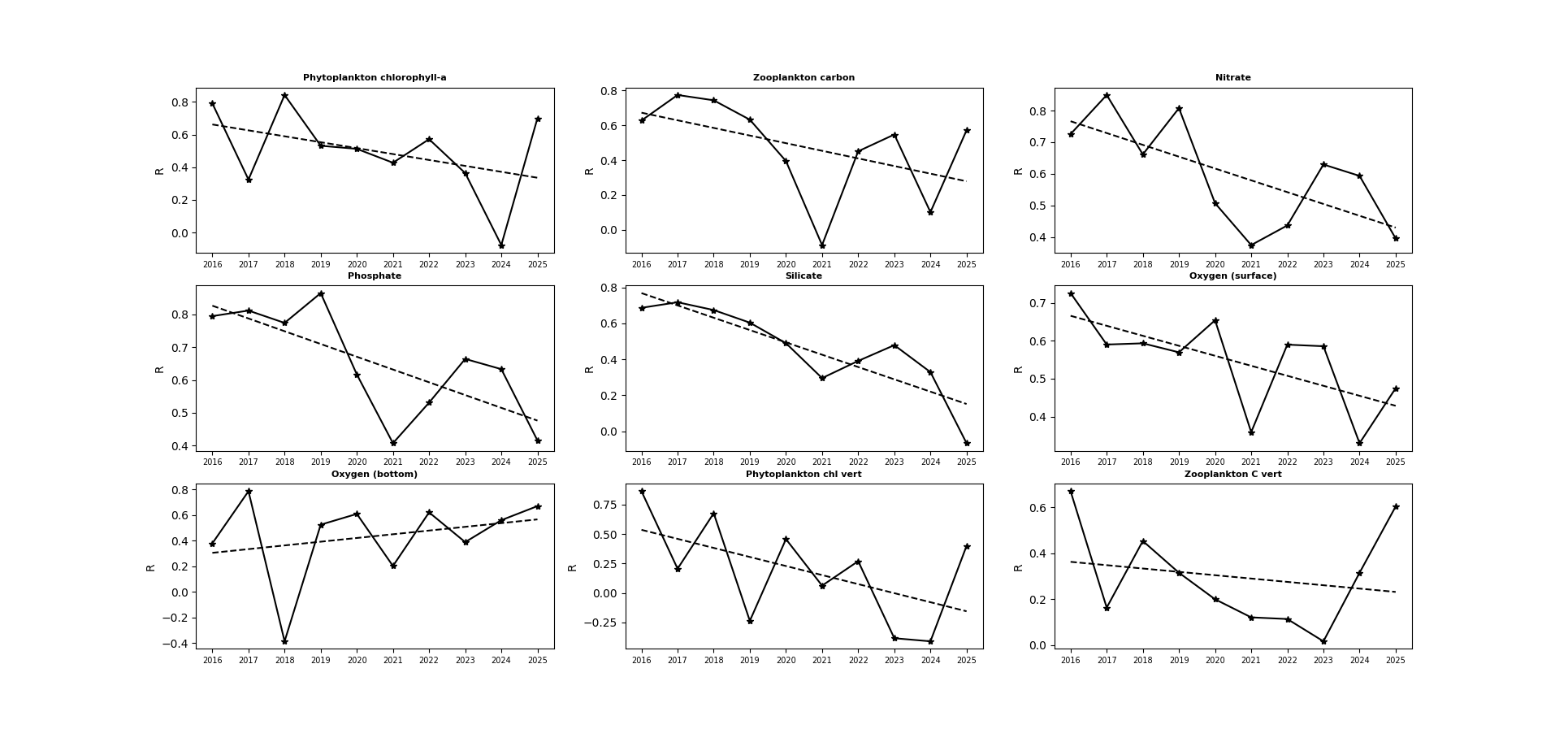}
    \caption{The Pearson correlation calculated for the LSTM ensemble median and validation-test data separately for each year throughout the validation and test period (2016-2025). The Pearson correlation is calculated for the selected biogeochemistry variables.}
    \label{fig:7}
\end{figure}

%4. How well can the emulator forecast anomalies on the decadal scale? Here anomalies are for both, emulators and the validation-test data calculated from the training data climatology (since we are interested in the departures from known behaviors, which is the climatology from the GOTM-FABM-ERSEM model run). This question is answered by Fig.\ref{fig:8} and Fig.\ref{fig:A4} of the Appendix. The results are more mixed.

\begin{figure}[h!]
    \hspace{-2.5cm}
    \includegraphics[width=18cm, height=9cm]{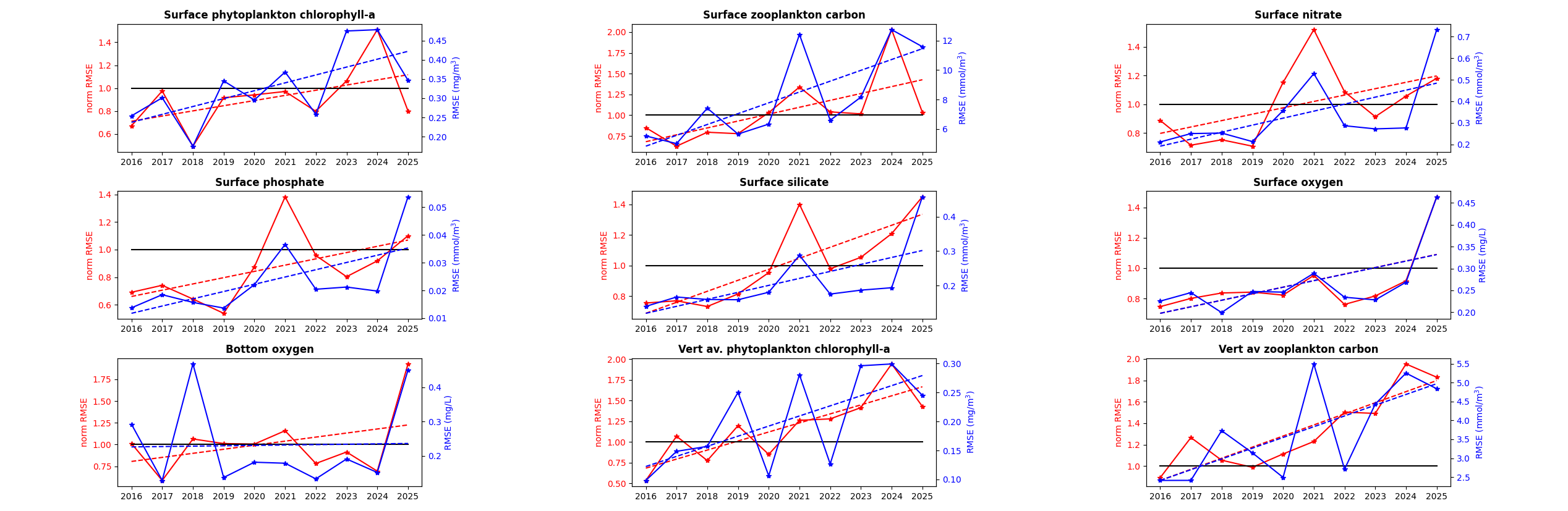}
    \caption{The LSTM ensemble median RMSE relative to the reference ERSEM simulation for each year of the validation-test data period. The same panels show LSTM RMSE normalized to the RMSE of a zero-anomaly prediction model (no knowledge of anomalies, all future prediction is climatology).}
    \label{fig:8}
\end{figure}

We investigated the stability of the decadal LSTM forecast relative to a red, skewed, multiplicative noise applied to its inputs. The multiplicative noise was in the most extreme case drawn (independently for each variable) as a white noise from a uniform distribution across the [0.1, 3.5] interval and then, to introduce temporal autocorrelations, it was low-pass filtered with a Gaussian filter with standard deviation of 15 days. For the physics-informed 1D CNN emulator, the same multiplicative noise was applied across the vertical water column. The impact of the skewed noise on the physics inputs to the emulator was significant, at least for the most extreme noise explored, as shown in Fig.\ref{fig:A5}:A in the Appendix. However the impact of this noise on the LSTM emulator was comparably smaller, as shown in Fig.\ref{fig:A5}:B of the Appendix. This means that the LSTM emulator, once trained, is highly stable with respect to the uncertainty and biases of its physical inputs. This stability could be important when the emulator is driven by inputs from a physics emulator, observations, or reanalysis data, rather than by the process-based simulator on which it was trained. Unlike the LSTM emulator, the 1D CNN emulator is much less stable, i.e. when driven by the physics inputs with the same correlated skewed noise from Fig.\ref{fig:A5}:A, the emulator quickly diverged from a sensible trajectory (not shown here). 
Similar sensitivity of the 1D CNN was observed when key hyperparameters, such as the 10-day training rollout window, were modified, or when changes to the network architecture were introduced that otherwise improved the validation score (evaluating short-term, 10-day, model forecast). 
%Similar sensitivity of the 1D CNN was observed when some key hyperparameters, such as the 10-day training rollout window were changed, or updates to the network architecture were made that improved network validation score (i.e. 10-day short-term dynamics). 
%which could be relevant if the emulator was run on physics emulator inputs (or physics observations, or reanalysis), rather than a physics process-based simulator that it was trained on.  

%5. Discuss the stability relative to noise in inputs. LSTM is remarkably stable (see Fig.\ref{fig:A5} of the Appendix), 1D CNN needs testing {\bf (to do)}..

%6. Discuss the overall lower performance (and possibly lesser stability) of the 1D CNN - can we speculate that this has something with misrepresenting the vertical water column (too low MLD from Fig.\ref{fig:5}?)?

\subsection{Short-medium range prediction}

\begin{figure}[h!]
    \hspace{-2.5cm}
    \includegraphics[width=18cm, height=11cm]{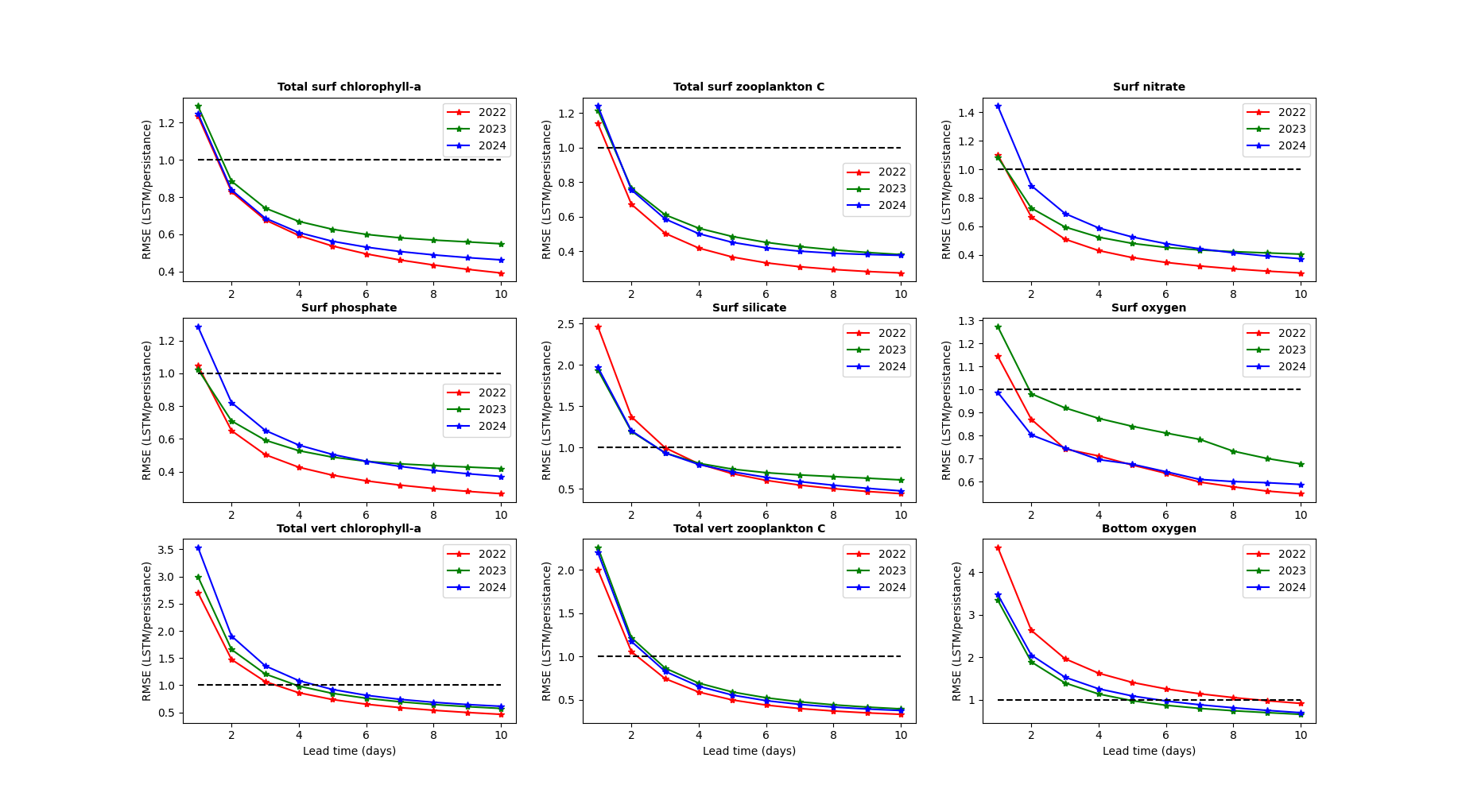}
    \caption{The RMSE skill of short-range (up to 10-day) prediction of LSTM emulator (ensemble median) relative to the persistence RMSE skill. The skill is compared across the 1-10 day forecast lead time. The skill was averaged across 3 separate years (from the 2022-2024 period). The 2025 year was left out due to the degradation of model skill in that year.}
    \label{fig:9}
\end{figure}

The LSTM emulator, trained on the GOTM-FABM-ERSEM free run simulation, is highly skilled in the short-medium range prediction when compared to the persistence metric (Fig.\ref{fig:9}). All the emulated variables except for sea bottom oxygen perform 50-60\% better in RMSE at the end of the 10-day forecasting period. The persistence is outperformed for most of the variables at the second forecasting day, and for all the variables except for the bottom oxygen at the fourth forecasting day (Fig.\ref{fig:9}). The results shown in Fig.\ref{fig:9} are in line with the study of \citet{smith2026identifying}, which focused on the Black Sea, and they support the idea that reduced complexity emulators (such as LSTM) are highly efficient tools for short-medium range forecasting. 

Unlike the LSTM emulator, the 1D CNN emulator performance in 10-day forecasting is typically worse than persistence across the full 10-day forecast period (Fig.\ref{fig:A6}). There is a possibility that this is due to the larger difference in seasonal climatology between the 1D CNN emulator and the GOTM-FABM-ERSEM simulator (Fig.\ref{fig:6}). It is quite possible that during the forecast period the 1D CNN diverges to its own climatological dynamical trajectory, degrading its skill relative to the persistence, which originates from the simulator run. Similar behaviour was discussed by \citet{skakala2018assimilation}, in the context of the UK regional three-dimensional model chlorophyll-$a$ forecast, where the 5-day model forecast skill is outperformed by the persistence, due to the differences between the model simulated chlorophyll-$a$ seasonal climatology and the observational seasonal climatology. Interestingly, we also explored a modified 1D CNN architecture incorporating dropout regularization and a cross-depth dense bottleneck. This substantially improved the model's validation performance and short-term forecasts of vertically averaged variables (results not shown). However, the architecture proved unstable during long autoregressive rollouts. Consequently, the 1D CNN types-of emulator may be better suited to short-term forecasting applications, which are arguably a more natural use case for models of this complexity.

%Interestingly, we explored a 1D CNN architecture where we introduced dropout regularization and a cross-depth dense bottleneck. This substantially improved the validation score of the 1D CNN model and the short-term forecast of the vertically averaged variables (not shown here). This architecture however has proven to be unstable on long rollout times. The 1D CNN type-of emulator could be therefore generally considered for short-term forecasting, which is likely a more natural application of such high complexity models.  

%{\bf Why CNN is so poor? Is that because it has larger climatological biases as can be seen in Fig.\ref{fig:6}? Can this be (indirectly) inferred?}

\subsection{Outperforming the forecasting system}

\begin{figure}[h!]
    \hspace{-1.5cm}
    \includegraphics[width=16cm, height=11cm]{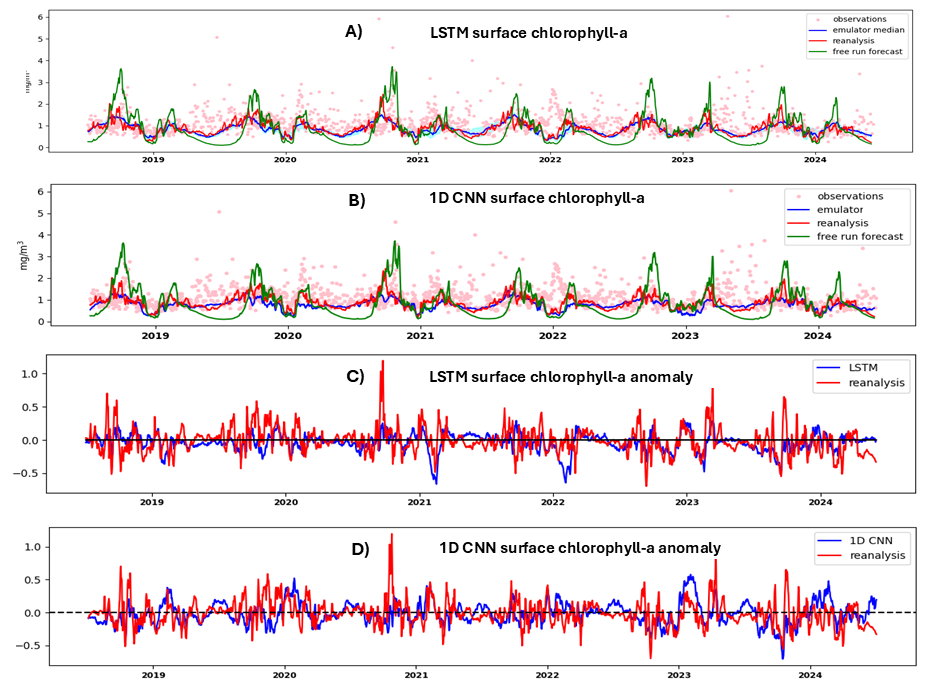}
    \caption{The upper two panels (one for LSTM, another for 1D CNN) compare the surface total chlorophyll-$a$ concentrations for the reanalysis (red), the emulator (blue), the GOTM-FABM-ERSEM simulator forecast (green) and the assimilated OC-CCI observations (pink dots). For LSTM the blue represents ensemble median and the aqua colour the two quartiles around median. Prediction of anomalies is shown for the emulators in the two rows at the bottom of the Figure. The time-series are compared for the test period (2019-2025). The GOTM-FABM-ERSEM forecast (free run) was initialized from the reanalysis on the 1 January 2019, similarly to the emulators.}
    \label{fig:10}
\end{figure}

\begin{figure}[h!]
    \hspace{-2cm}
    \includegraphics[width=17cm, height=12cm]{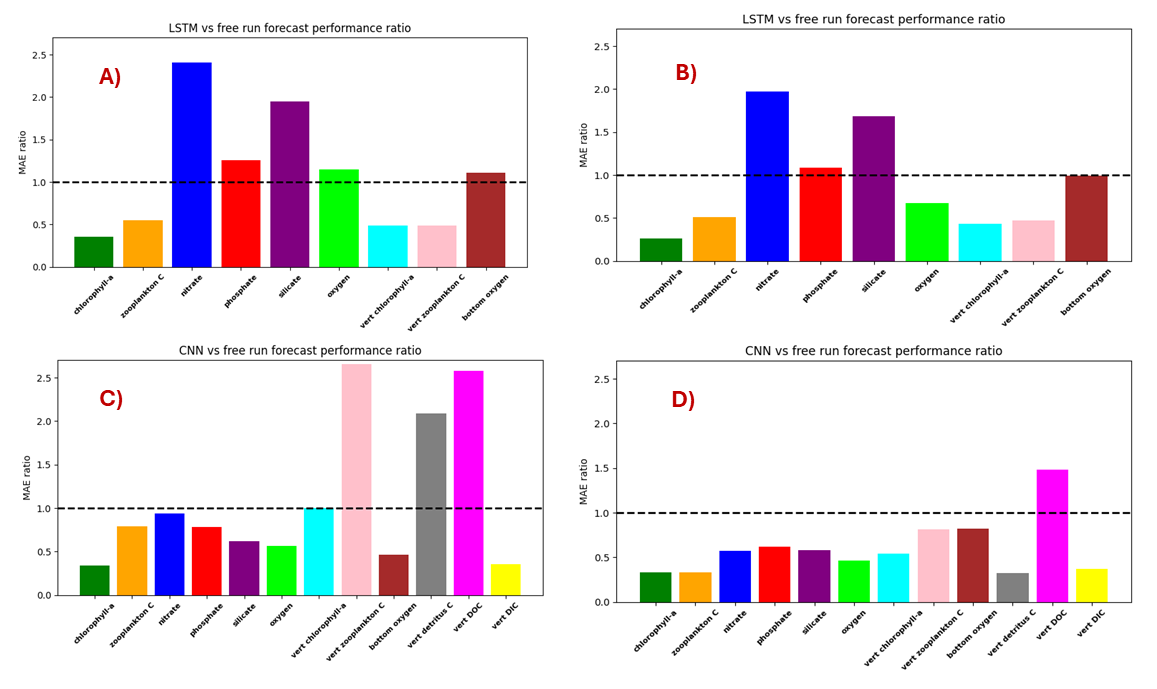}
    \caption{Left-hand panels A, C show the MAE skill of the emulators to predict the reanalysis data across the test period (2019-2025) relative to the same MAE skill of the free run. The right-hand panels B, D show the same for low-pass filtered data with a 21 day window. This is to demonstrate the impact of hard-to-predict assimilation cycle on the emulator skill to predict anomalies. The upper panels A-B show the LSTM ensemble median and the bottom panels C-D the 1D CNN emulator.}
    \label{fig:11}
\end{figure}

When trained on the reanalysis, both LSTM and 1D CNN substantially outperform the 7-year model free run simulator forecast (Fig.\ref{fig:10}). The simulator forecast and the two emulator runs from Fig.\ref{fig:10} were all started from the reanalysis on the 1\textsuperscript{st} January 2019. As can be seen in Fig.\ref{fig:10}:A-B, the assimilation of satellite chlorophyll-$a$ has major impact on the surface chlorophyll-$a$ time-series, removing the free ERSEM run systematic seasonal biases by increasing the Winter surface chlorophyll-$a$ concentrations and reducing the magnitude of Spring phytoplankton bloom. These biases are well documented for ERSEM three-dimensional applications in the North-West European Shelf (NWES) seas, and similar corrections to those biases are performed, using data assimilation, in NWES reanalysis and forecasting systems \citep[e.g.,][]{skakala2018assimilation, skakala2022impact, fowler2023validating}. The deep learning emulators trained on the reanalysis and run for the 2019-2025 7-year forecast are inherently correcting the ERSEM seasonal biases, and therefore produce much more skilled forecast than the free run simulator, when compared to the assimilated satellite observations, or the reanalysis (Fig.\ref{fig:10}:A-B and Fig.\ref{fig:11}:A,C). This is true for most of the predicted variables (Fig.\ref{fig:11}:A,C). However, the emulators are not as good in forecasting the daily anomalies within the reanalysis (Fig.\ref{fig:10}:C-D, with Pearson correlation, $R$, around 0.4). This can be explained by the fact that these anomalies are strongly affected by data assimilation, which is to a degree inherently unpredictable, because the assimilation cycle depends on the intermittent satellite data availability, while the analysis state is also affected by uncertainties in the observational data. Low-pass filtering of both the forecasts and the reanalysis can remove much of the randomness introduced by data assimilation, which may then result in improved emulator skill relative to the free-run simulator forecast (e.g. see Fig.\ref{fig:11}:B). 

%Can we outperform the model forecast by training the emulator with reanalysis, so it accounts for inherent model biases? This is shown both for LSTM and 1D CNN in Fig.\ref{fig:9} and the answer is positive. However this is mainly due to good prediction of the reanalysis climatology rather than of anomalies, as can be also seen in Fig.\ref{fig:9}. Fig.\ref{fig:10} shows for LSTM wider evaluation of the model skill in predicting reanalysis compared to the GOTM-FABM-ERSEM forecast.

%\textcolor{red}{{\bf 
%Why there is little skill to predict anomalies? I guess this is due to DA and observational uncertainty, which is largely unpredictable from emulator point of view. However how does prediction improve if we take low-pass filtered anomalies? 
%Fig.\ref{fig:10} needs updating with barplot showing skill on low-pass filtered data and also 1D CNN skill. Should be 2 x 2 Figure.}}

%\clearpage
\subsection{Explainable AI Sensitivity Analysis}
%{\bf \textcolor{red}{JZ version}}
\begin{figure}[h!]
    \centering
    \includegraphics[width=1\linewidth]{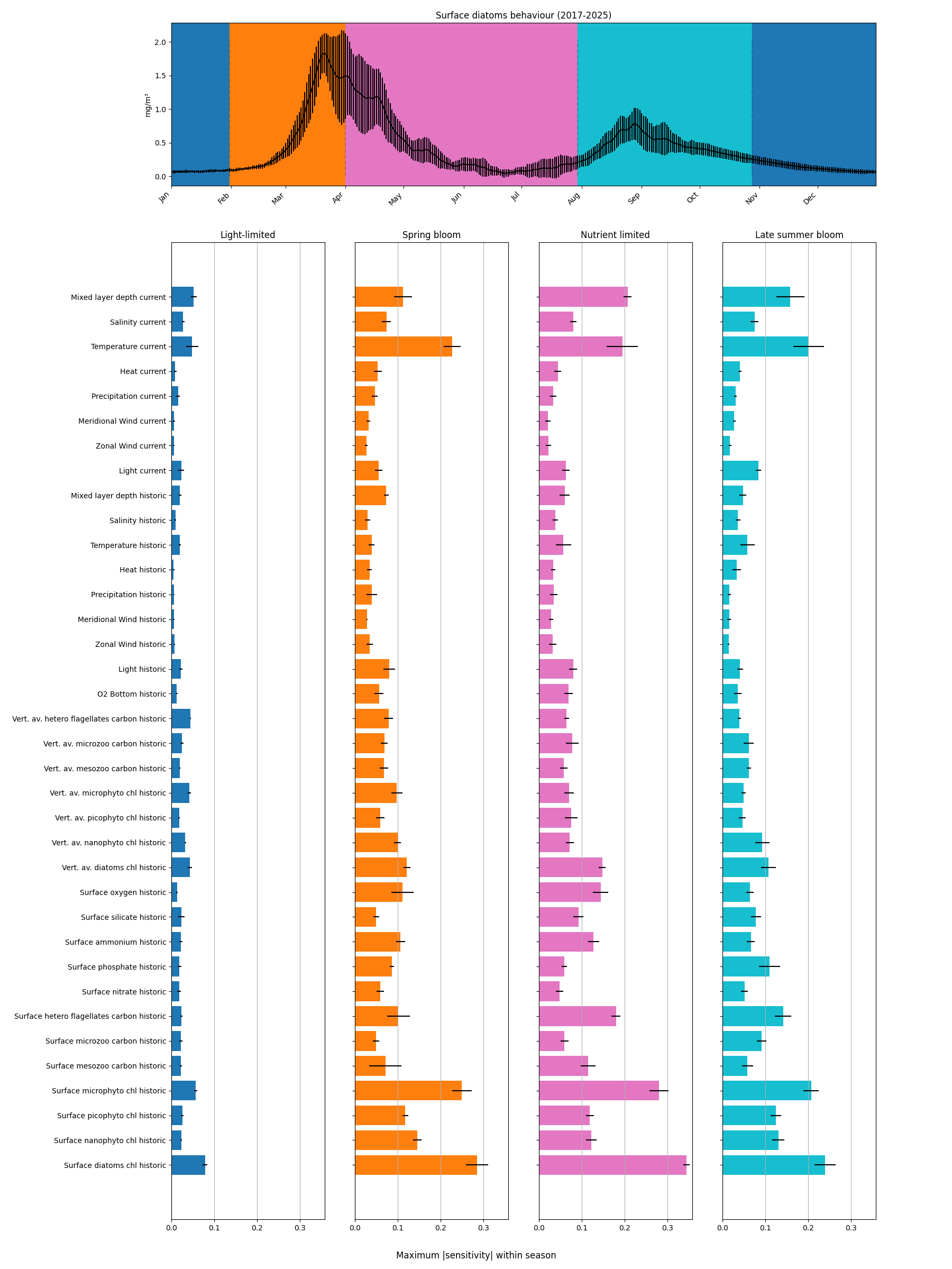}
    \caption{ The top panel shows the daily mean $\pm1$
     standard deviation of surface diatom chlorophyll over the 2017-2025 validation and test period; shaded regions denote key seasonal regimes of phytoplankton behaviour. The remaining panels show, for each seasonal regime, the mean $\pm1$ standard deviation across years of the maximum absolute sensitivity of surface diatom chlorophyll to each input. Seasonal regimes windows are fixed across years.}
    \label{fig:seasonal_summary}
\end{figure}

\begin{figure}[h!]
    \vspace{-1cm}
    \hspace{-1cm}
    \centering
    \includegraphics[width=1\linewidth]{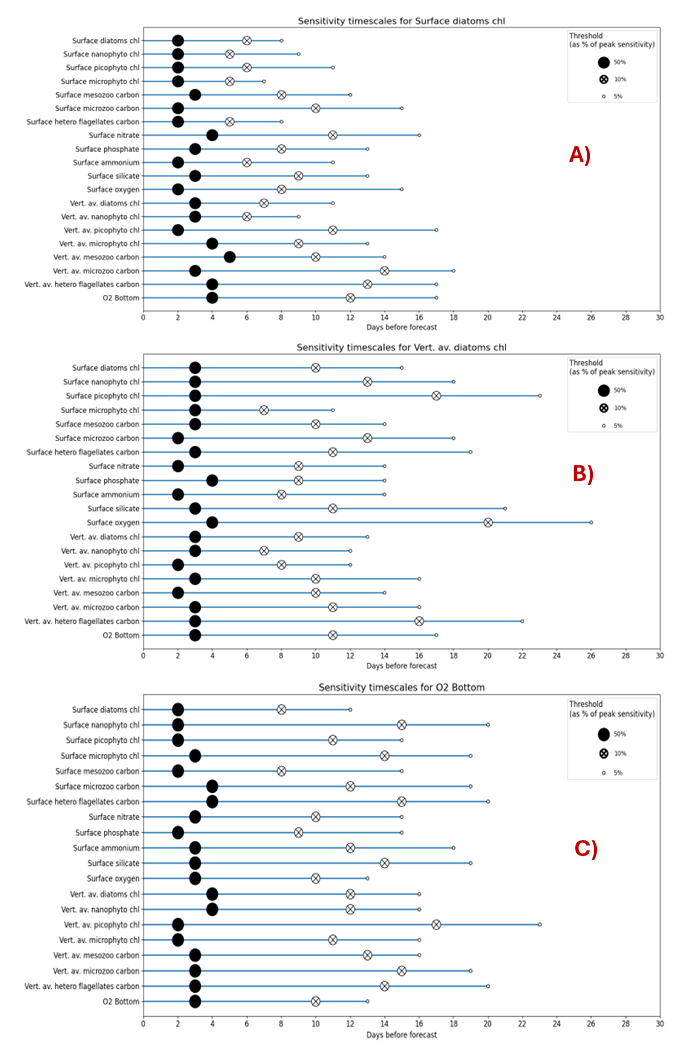}
    \caption{Mean lag at which sensitivity of surface diatoms chlorophyll-$a$ (A), vertically averaged diatoms chlorophyll-$a$ (B), and sea bottom dissolved oxygen concentration (C) to each forcing input first drops below 50\% (filled circle), 10\% (circle with cross), and 5\% (small open circle) of its peak mean value, summarising the effective memory timescales across the inputs.}
    \label{fig:xai_timescale_bio}
\end{figure}

\begin{figure}[h!]
    \centering
    \includegraphics[width=1\linewidth]{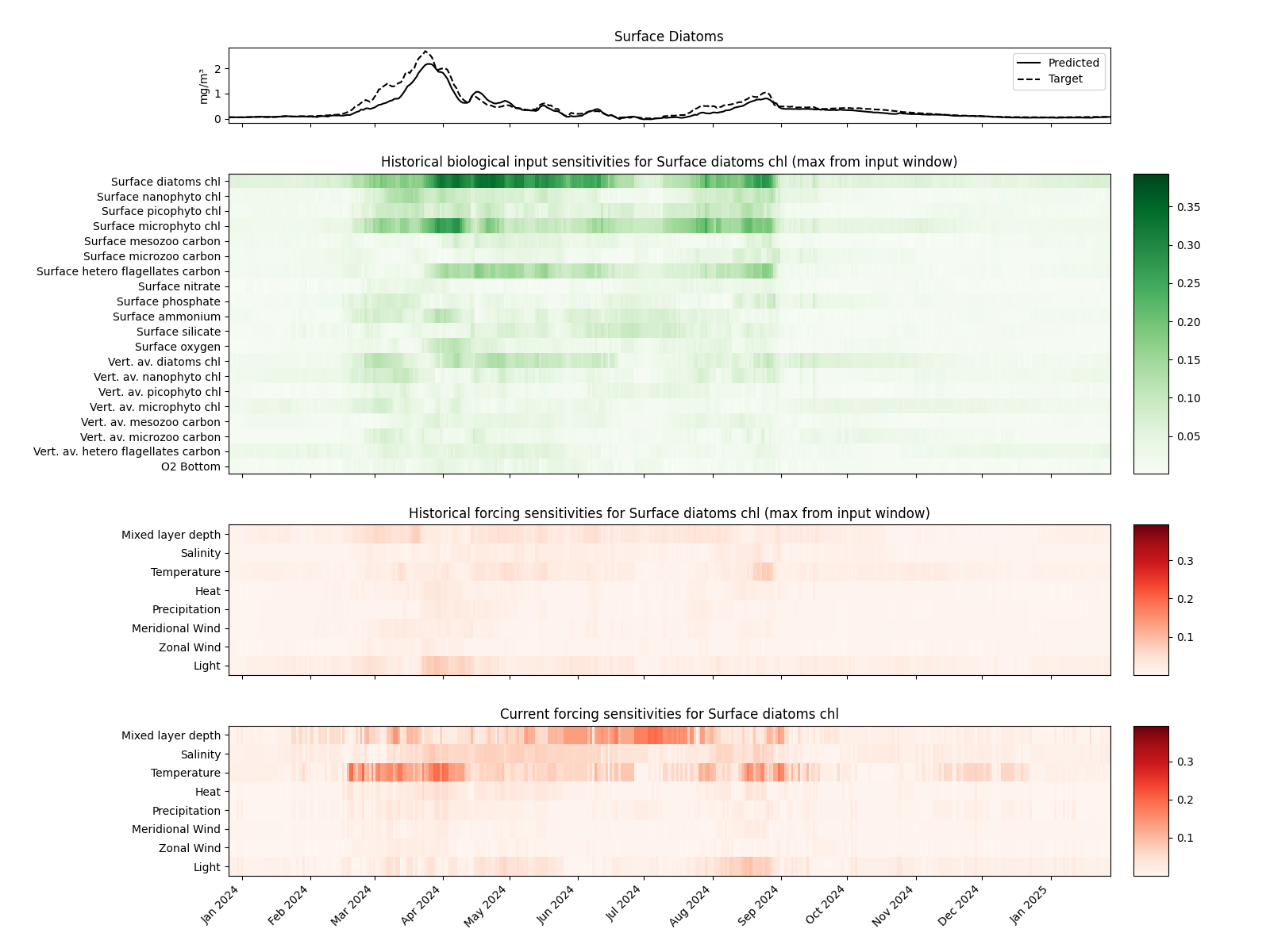}
    \caption{ Top panel shows predicted and target value of surface diatoms chlorophyll over 2024-2025.
    Remaining panels show temporal evolution of absolute sensitivity of surface diatoms chlorophyll to each biological/forcing input, taking the maximum sensitivity within the 30-day lookback window for historic variables. Darker colours indicate periods of stronger sensitivity. Zoomed into 2024-2025. }
    \label{fig:xai_single_year}
\end{figure}

We applied explainability analyses to the LSTM model, as it proved to be the better performing of the two approaches tested. In addition, its lower complexity makes it more amenable to interpretation using explainability methods. To improve the interpretability of the model, we apply saliency analysis via backpropagation through the network, computing the sensitivity of each predicted output to each input variable across both the 30-day lookback window of historical state variables and the concurrent meteorological forcing prescribed for the forecast step. Among the predicted outputs, we focus here on three variables of the highest interest: the surface diatom chlorophyll-$a$ concentrations, the vertically averaged chlorophyll-$a$ and the sea bottom oxygen. The diatoms are the dominant phytoplankton type at L4 (not shown here) and their chlorophyll-$a$ concentration is a proxy for biological productivity. The sea bottom oxygen concentration is an indicator for oxygen deficiency (which is not typically an issue at the L4 station, but nonetheless interesting to understand from a process-level perspective). 

The biogeochemical state during the preceding 30 days is a strong predictor of the present state (Fig.\ref{fig:seasonal_summary}). This is not surprising, as biogeochemical variables typically evolve relatively slowly on daily timescales; indeed, a persistence forecast exhibits considerable skill at this temporal resolution (Fig.\ref{fig:9}). As expected, values from the previous day provide the most important, first-order, predictive information, with longer time-lag scales providing a higher-order information to which the prediction is less sensitive (Fig.\ref{fig:xai_timescale_bio}). 
%Fig.\ref{} indicate that the typical scale of memory needed from the inputs is about one week for the surface diatom chlorophyll-$a$ and closer to two weeks for the vertically averaged diatom chlorophyll-$a$ and the bottom oxygen. This is expected as the vertically averaged and the sea bottom variables have slower temporal evolution than the surface variables.     
Fig.\ref{fig:xai_timescale_bio} indicates that the effective memory timescale required by the model is approximately one week for surface diatom chlorophyll-$a$, and closer to two weeks for vertically averaged diatom chlorophyll-$a$ and bottom oxygen. This is consistent with the slower temporal variability of vertically averaged and bottom-water properties compared with surface variables. The highest sensitivity of surface diatoms is to their own historical values (Fig.\ref{fig:seasonal_summary}), which is expected from the autocorrelation time-scales of ERSEM variables. The second most sensitive predictor of diatoms identified by the LSTM is microphytoplankton, presumably because of the functional similarities between these groups. Their concentrations may be influenced by similar environmental drivers, and they may also compete for the same resources. Fig.\ref{fig:seasonal_summary} may provide further insights into the importance of biogechemistry predictors, but the relative sensitivities of these biogeochemical predictors should be interpreted with caution, as biogeochemical variables are often highly correlated. 

Recovering the seasonal climatology and achieving skill in predicting anomalies on decadal timescales requires adding physical (atmosphere, ocean) forcing variables to the past biogeochemical state. Fig. \ref{fig:seasonal_summary} and Fig.\ref{fig:xai_single_year} show that the most influential physical inputs for predicting surface diatom chlorophyll-$a$ are sea surface temperature (SST) and mixed-layer depth. SST is likely used by the LSTM as a proxy for seasonality, enabling the model to reproduce the seasonal climatology of the simulator. However, SST alone is unlikely to provide sufficient information to predict the inter-annual variability in bloom timing (Fig.\ref{fig:6.5}), as it does not directly capture vertical water column dynamics. This information is more likely encoded through the mixed-layer depth. The prominence of mixed-layer depth raises the question of whether the model's behaviour can provide insight into the mechanisms controlling diatom bloom formation at L4 in the GOTM–FABM–ERSEM simulation. Several hypotheses have been proposed to explain bloom initiation in temperate seas, including the critical depth hypothesis \citep{sverdrup1953conditions}, the critical turbulence hypothesis \citep{huisman1999critical}, and mechanisms emphasising grazer control \citep[e.g.,][]{behrenfeld2010abandoning}. In the present simulation, zooplankton grazing appears to play a limited role in bloom initiation. The time series of surface variables indicate that grazer biomass increases only after a substantial lag relative to phytoplankton growth (not shown), suggesting that grazing is unlikely to control the timing of bloom onset. Furthermore, the relatively high importance assigned by the LSTM to mixed-layer depth, compared with atmospheric forcing variables, may indicate that bloom timing is more strongly linked to water column structure than to short-term variations in atmospheric forcing. While this interpretation should be treated with caution, it is arguably more consistent with the critical depth framework than with the critical turbulence hypothesis.

%Finally, as predicting the sea bottom oxygen is of general interest to manage the risks of hypoxia, we have looked at the LSTM predictors of the bottom oxygen variable (see Fig.\ref{fig:xai_hov_o2bottom}). These have much more straightforward structure than the LSTM predictors of surface diatoms. The long autocorrelation time-scales of the slower dynamics at the sea bottom show in past sea bottom oxygen being the best predictor of its present values. Another predictor used by LSTM is temperature to capture the seasonality in the bottom oxygen values. This is followed by the mixed layer depth, which accounts for the connectivity between the sea bottom (at 50m) and the ocean surface, providing ventilation with the atmosphere.

Finally, because predicting bottom-water oxygen is important for assessing and managing the risk of hypoxia, we examined the LSTM predictors of bottom oxygen (Fig.\ref{fig:xai_hov_o2bottom} of Appendix). The predictor structure is considerably more straightforward than that for surface diatoms. The relatively slow dynamics of the deeper waters result in long autocorrelation timescales, making past bottom-oxygen concentrations the strongest predictor of present values (Fig.\ref{fig:xai_hov_o2bottom}). Temperature is another important predictor, likely reflecting the seasonal variability of bottom oxygen. This is followed by mixed-layer depth, which captures the degree of connectivity between the seabed (at 50 m depth) and the ocean surface, thereby influencing ventilation from the atmosphere.

%The sensitivity structure in Fig.\ref{fig:xai_hov_diatoms} for surface diatom chlorophyll-$a$ provides nice independent insights into phytoplankton dynamics at L4. Firstly, the diatom surface concentration is strongly sensitivity to its past values, as well as the past values of other surface PFTs chlorophyll-$a$. Strong self-sensitivity is expected given the autocorrelation present in most biogeochemical variables. The phytoplankton dynamics at L4 is strongly driven by the bloom, which according to the 

%\textcolor{red}{{\bf 1. key question - I would dive into the onset of Spring bloom: critical depth, critical turbulence, or zooplankton grazing? 2. we can discuss drivers of sea bottom oxygen, but here the analysis is limited, 3. we shall discuss the autocorrelation time-scales and how they differ among surface, vert averaged variables and the bottom oxygen.}}

\section{Conclusions}

In this study, we investigated the feasibility of using deep learning to emulate a highly complex regional marine biogeochemical model and to couple these emulators with an ocean physics simulator for biogeochemical forecasting across a range of time scales. Two emulator architectures were explored: (a) an LSTM neural network that emulates selected variables of interest, primarily at the surface, using a daily time step; and (b) a physics-informed 1D CNN emulator that reproduces the daily time-step operator of the biogeochemical model for all pelagic variables throughout the full water column. To enable a systematic exploration of emulator designs and a wide variety of research questions, we focused on a one-dimensional water column configuration at a station in the western English Channel.

Our results demonstrate that complex marine biogeochemical dynamics can be successfully emulated for forecasting applications ranging from weekly predictions to decadal time scales. The emulators also showed good skill in predicting anomalies, including LSTM predicting well the timing of phytoplankton blooms at the study location. These results are, however, all conditional on the availability of high-quality ocean physics forecasts, which are supplied as inputs to the biogeochemistry emulators. Furthermore, when trained on reanalysis data, the emulators were able to outperform the existing forecast generated by the numerical simulator. We also showed that the LSTM emulator remains stable in the presence of noise and biases in the ocean physics inputs, which is highly desirable if it was coupled to a separate physical ocean model emulator. Unlike the LSTM, the 1D CNN emulator was much more dependent on the quality of the simulated physics and showed in the presence of considerable anomalies signs of unstable behavior. Furthermore, the training process for the 1D CNN emulators was highly sensitive to the choice of random seed, converging to a range of solutions that, while capable of robust short- to medium-term predictions, exhibited markedly different behaviour during long-term emulator rollouts. Whether this can be improved upon will be explored in the future. Finally, we presented explainability analyses for several indicators of particular interest, demonstrating how these results could be used to gain potential insights into the mechanisms driving Spring bloom formation in the western English Channel.

Although extending this framework to three-dimensional configurations will inevitably introduce greater complexity, involving more sophisticated models and substantially higher computational costs for training, we expect the key conclusions of this study to remain valid. Exploring the emulators in fully three-dimensional systems represents an important direction for future research.

Looking ahead, we anticipate that deep learning emulators will become central to marine biogeochemical forecasting. They offer the potential to represent highly complex processes at high spatial resolution in coastal and regional systems while maintaining computational costs that are compatible with operational applications. Such approaches may also facilitate the routine use of ensemble forecasts and uncertainty quantification. A key remaining research challenge is the integration of these emulators with data assimilation techniques, enabling systematic correction of the ocean state within operational forecasting systems.

{\bf Acknowledgments:} JS and DM acknowledge financial support from the UK Natural Environment Research Council (NERC). JS is supported by the single centre national capability programme—Atlantic Climate and Environment Strategic Science (Atlantis), and National Centre for Earth Observation (NCEO). The authors are thankful to NERC Earth Observation Data Analysis and Artificial-Intelligence Service (NEODAAS) for supplying them with the satellite OC-CCI chlorophyll-$a$ observations assimilated in the reanalysis, and compute resource. The authors are also thankful to the Western Channel Observatory (WCO) service at PML for supplying them with L4 observations used to force the one-dimensional model, as well as to Helen Powley for downloading part of the ERA5 atmospheric forcing data for the same purpose. The code, the models and the key outputs from this study can be freely accessed on the GitHub repository https://\-github.com/\-JOZSKA/\-AI-1D-model-emulators-at-L4.

\clearpage
\bibliography{References}

\clearpage
\appendix

\section{Additional Figures}
\renewcommand{\thefigure}{A\arabic{figure}}

\begin{figure}[h!]
    \centering
    \includegraphics[width=\textwidth]{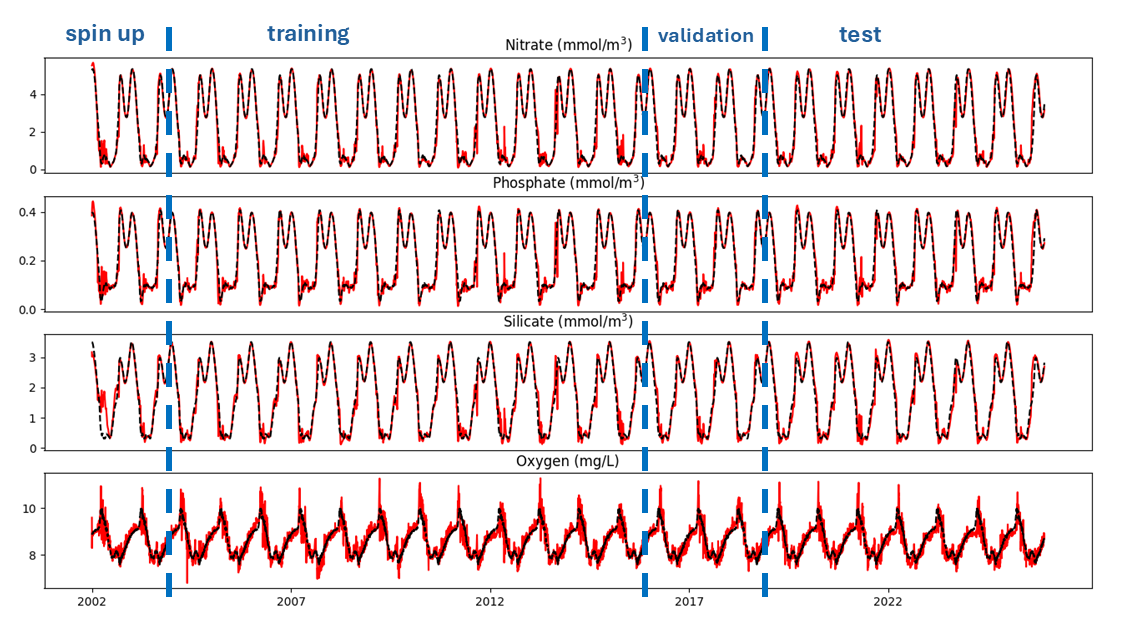}
    \caption{The ocean surface time series for four selected variables from the GOTM-FABM-ERSEM simulation spanning the 2002-2025 period. The different data (training/validation/test) are clearly marked by the vertical lines.}
    \label{fig:A1}
\end{figure}

\begin{figure}[h!]
    \centering
    \includegraphics[width=\textwidth]{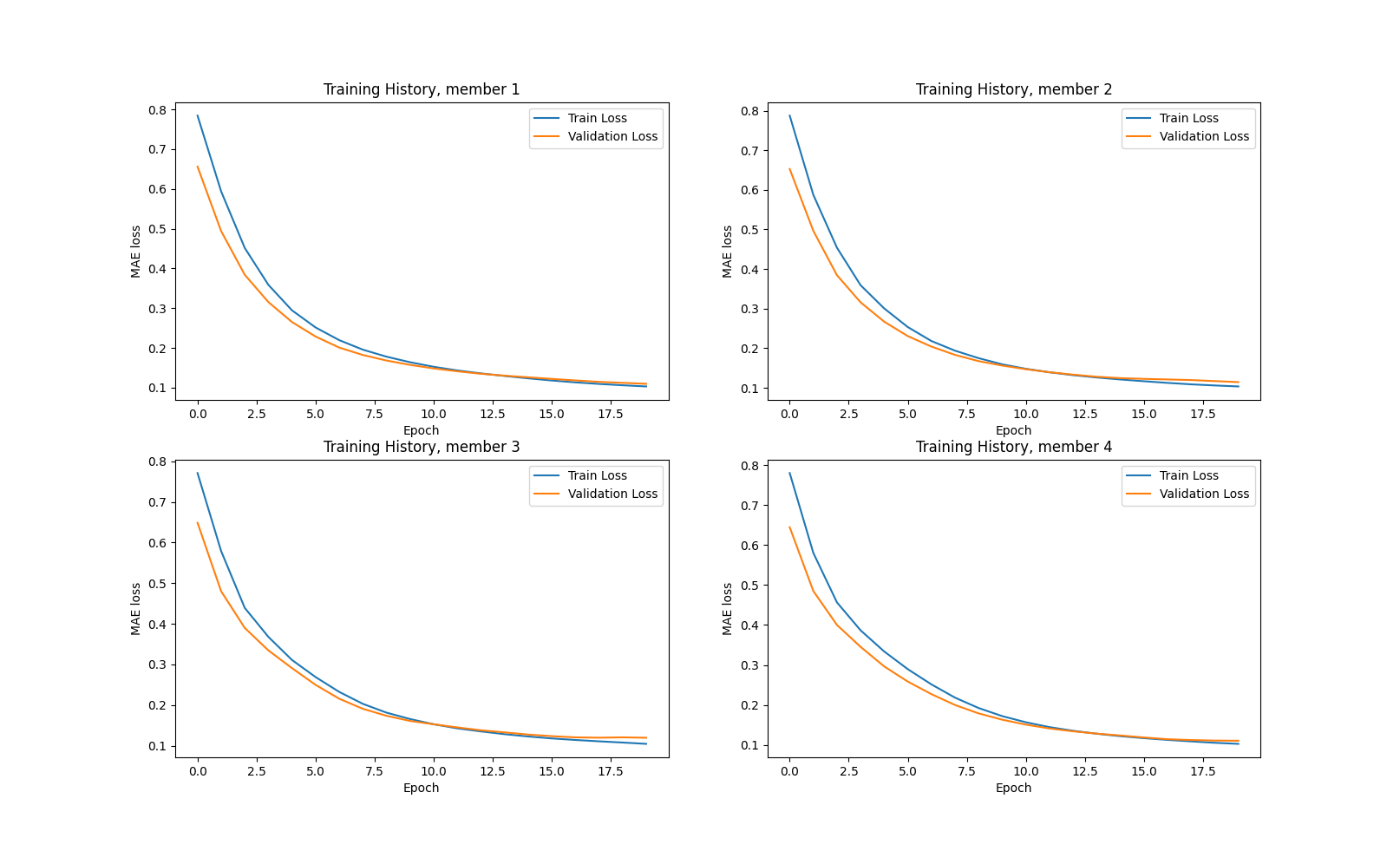}
    \caption{The training MAE loss function curves for LSTM and 4 randomly selected ensemble members.}
    \label{fig:A1.1}
\end{figure}

\begin{figure}[h!]
    \centering
    \includegraphics[width=\textwidth]{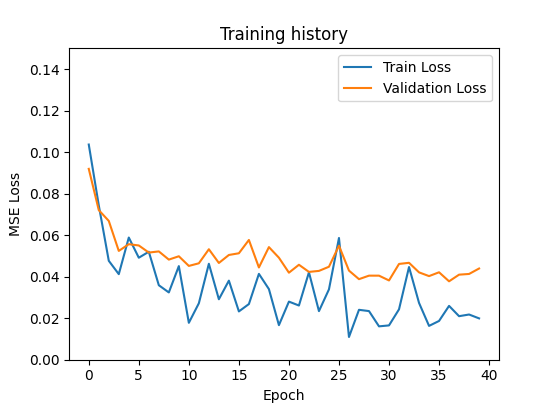}
    \caption{The training MAE loss function curves for the selected 1D CNN model.}
    \label{fig:A1.2}
\end{figure}

\begin{figure}[h!]
    \vspace{-1cm}
    \hspace{-2cm}
    \includegraphics[width=18cm, height=6.5cm]{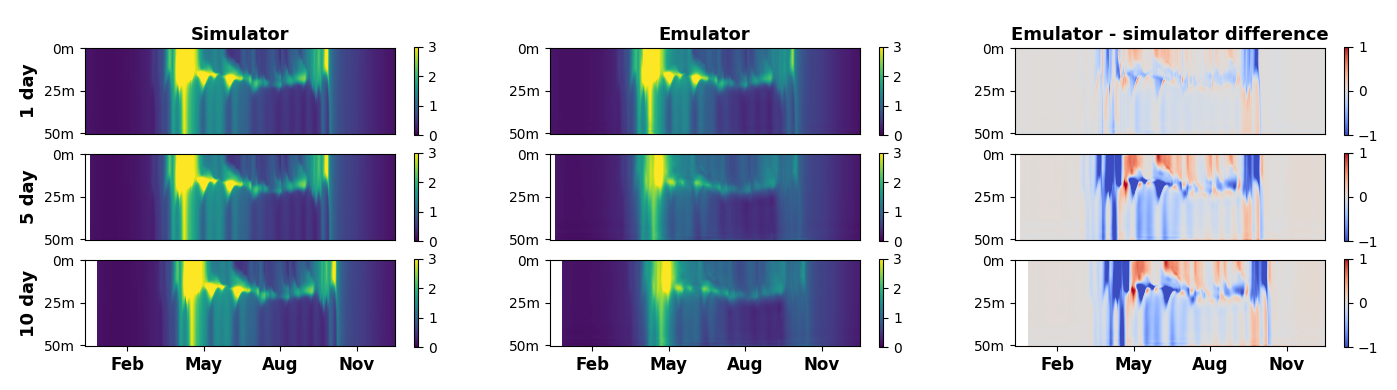}
    \caption{Hovm\"oller plots comparing year 2023 short-range 1D CNN emulator forecast with simulator data. The compared lead times are 1, 5 and 10 days. The variable compared is the total chlorophyll-$a$ concentration.}
    \label{fig:A2.5}
\end{figure}

%\begin{figure}[h!]
%    \vspace{-1cm}
%    \hspace{-2.5cm}
%    \includegraphics[width=18cm, height=20cm]{CNN_Hovmoller.png}
%    \caption{Hovm\"oller plots compared for eight selected variables across the test period (2019-2025) between the test data (left-hand panels) and the median of the 1D CNN emulator ensemble (middle column). The difference between those two (emulator minus test data) is shown in the right-hand panels. The eight different variables are plotted in the different rows.}
%    \label{fig:A2}
%\end{figure}

\begin{figure}[h!]
    \vspace{-1cm}
    \hspace{-2cm}
    \includegraphics[width=18cm]{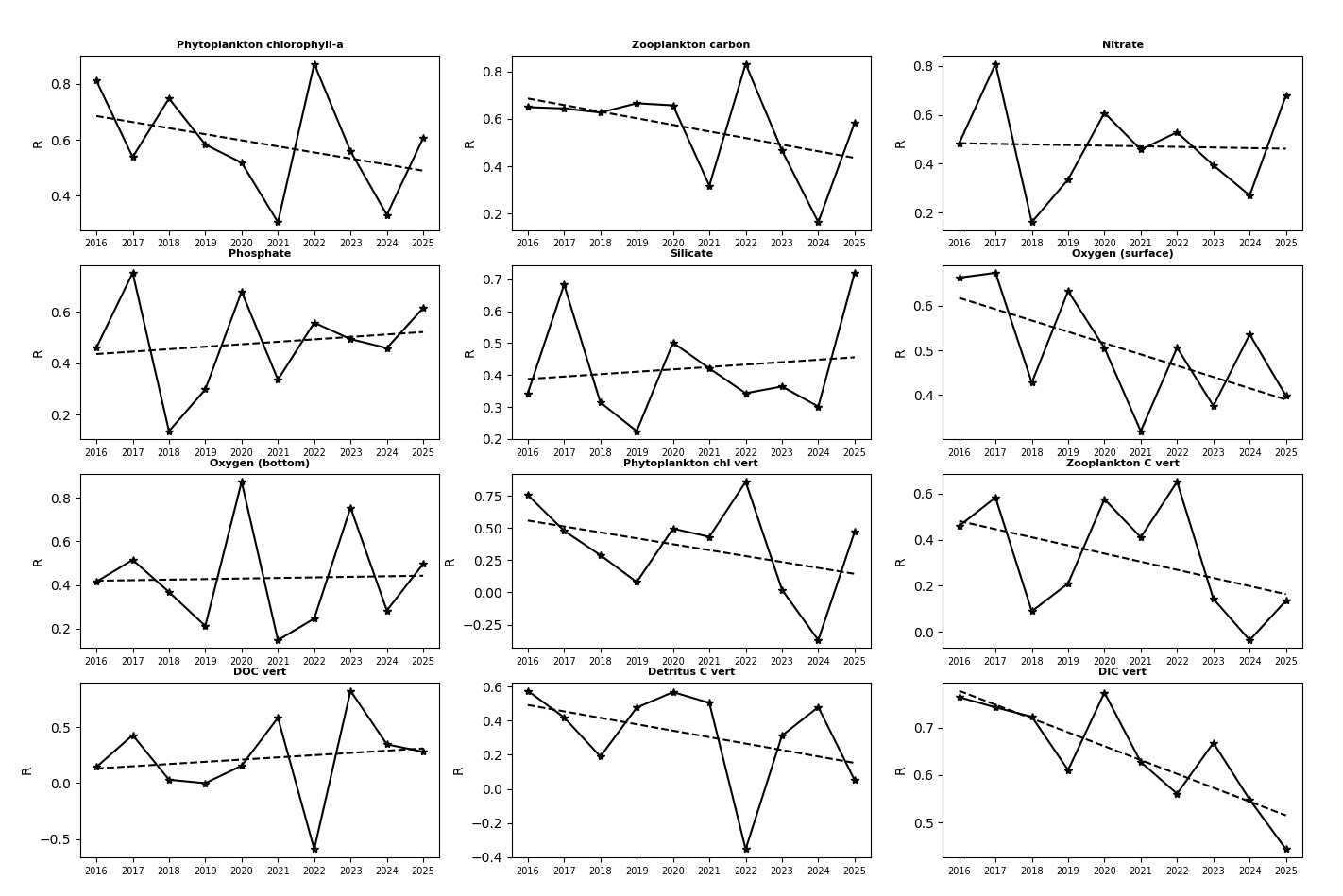}
    \caption{The same as in Fig.\ref{fig:6}, but for 1D CNN network.}
    \label{fig:A3}
\end{figure}

\begin{figure}[h!]
    \hspace{-2cm}
    \includegraphics[width=18cm, height=12cm]{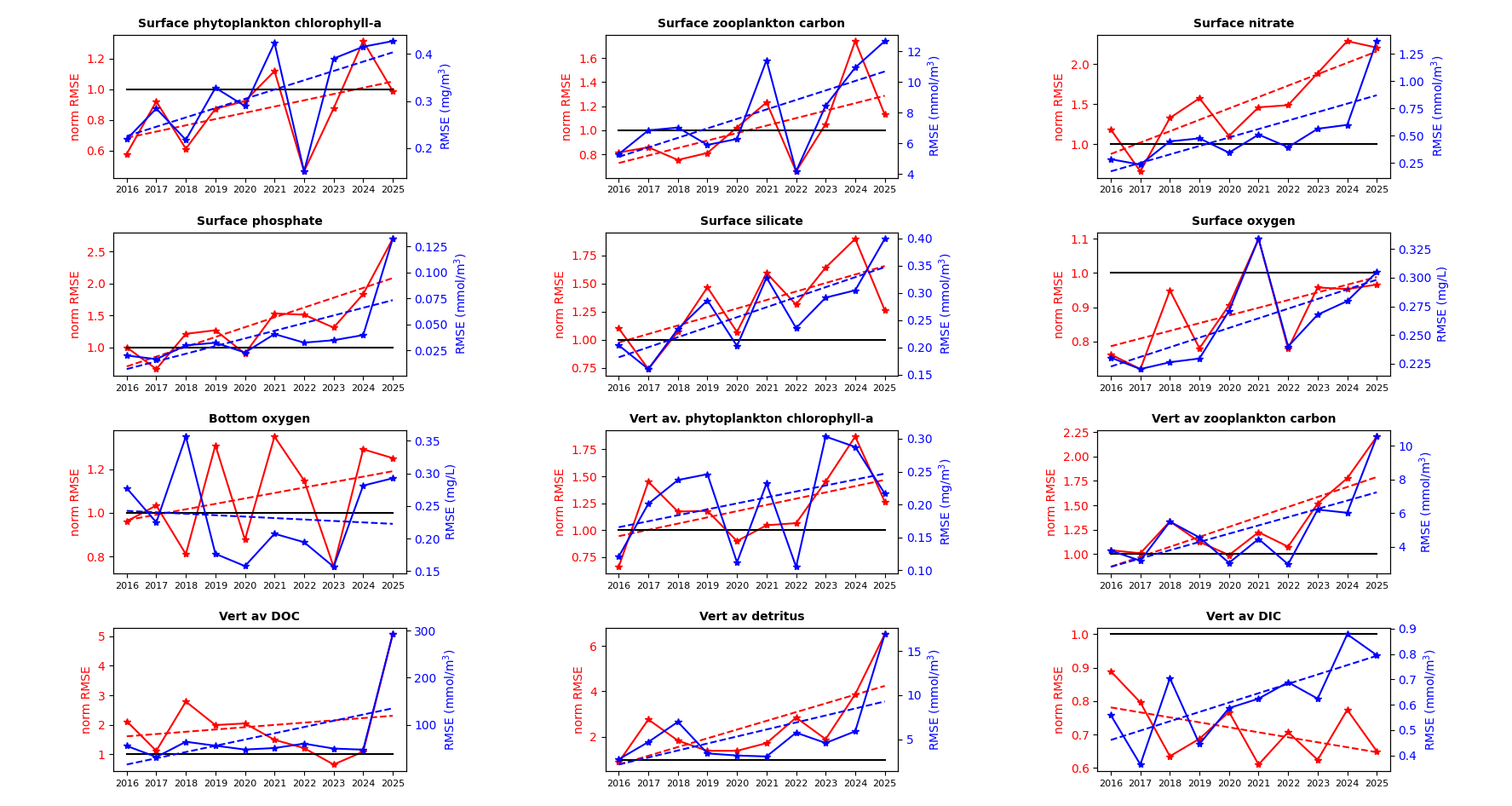}
    \caption{The same as in Fig.\ref{fig:7}, but for 1D CNN model.}
    \label{fig:A4}
\end{figure}

\begin{figure}[h!]
    \vspace{-1cm}
    \hspace{-2cm}
    \includegraphics[width=17cm, height=12cm]{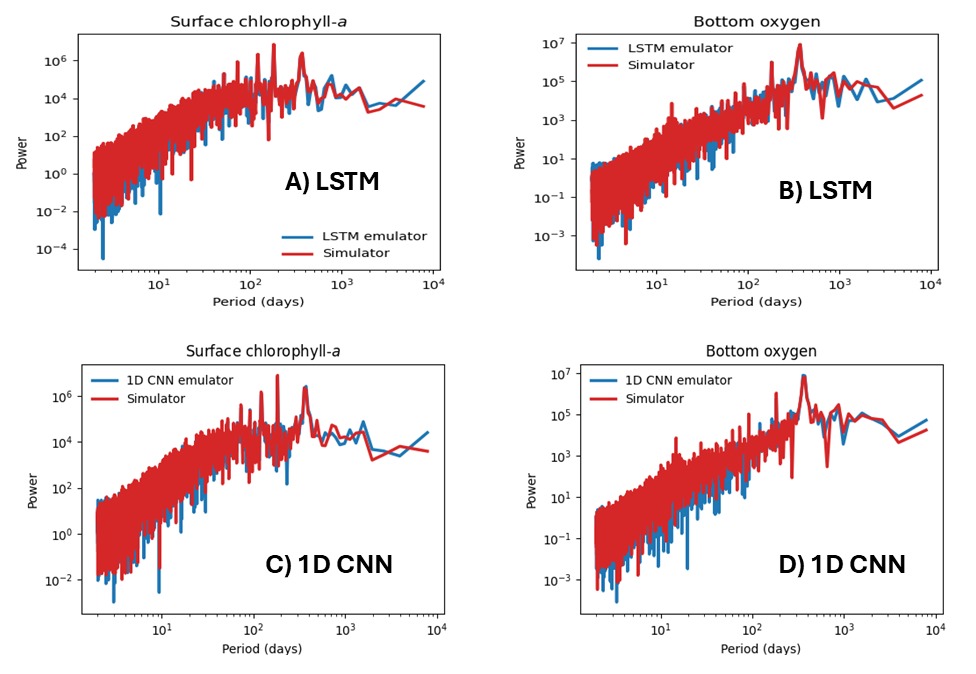}
    \caption{Comparing Fourier spectra between the GOTM-FABM-ERSEM simulator and the two emulators for two selected variables: surface total chlorophyll-$a$ and bottom oxygen. For LSTM a random ensemble member was selected to show the spectra.}
    \label{fig:AFourier}
\end{figure}

\begin{figure}[h!]
    \hspace{-1.5cm}
    \includegraphics[width=16cm, height=10cm]{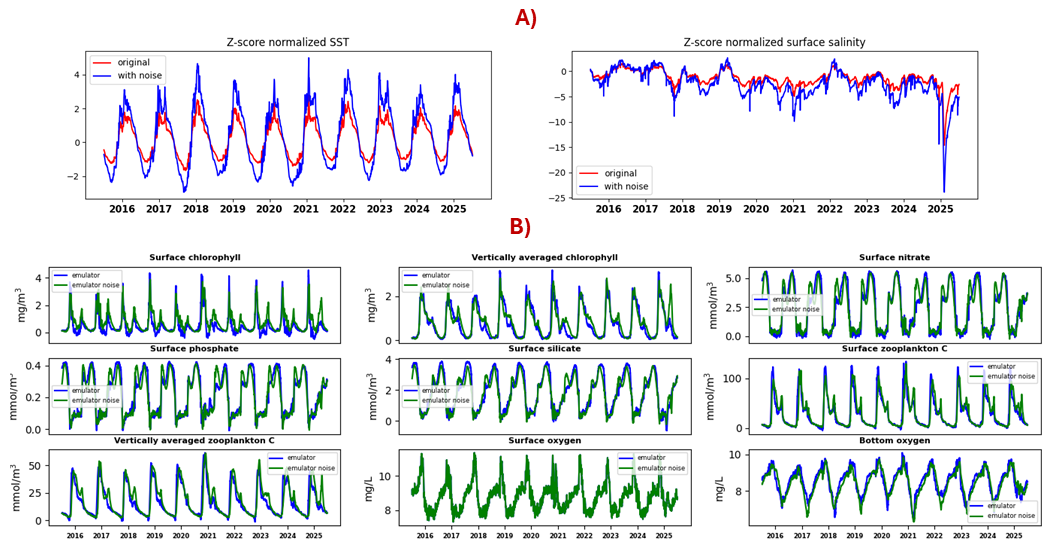}
    \caption{Impact of noise on the stability of the LSTM emulator. The panels in A show the impact of the red, skewed, noise on two selected physical ocean model inputs (SST and surface salinity) and the panels in B show the impact of the noise on the emulator forecast across the validation and test data period for a range of selected variables.}
    \label{fig:A5}
\end{figure}

\begin{figure}[h!]
    \hspace{-2cm}
    \includegraphics[width=16cm, height=12cm]{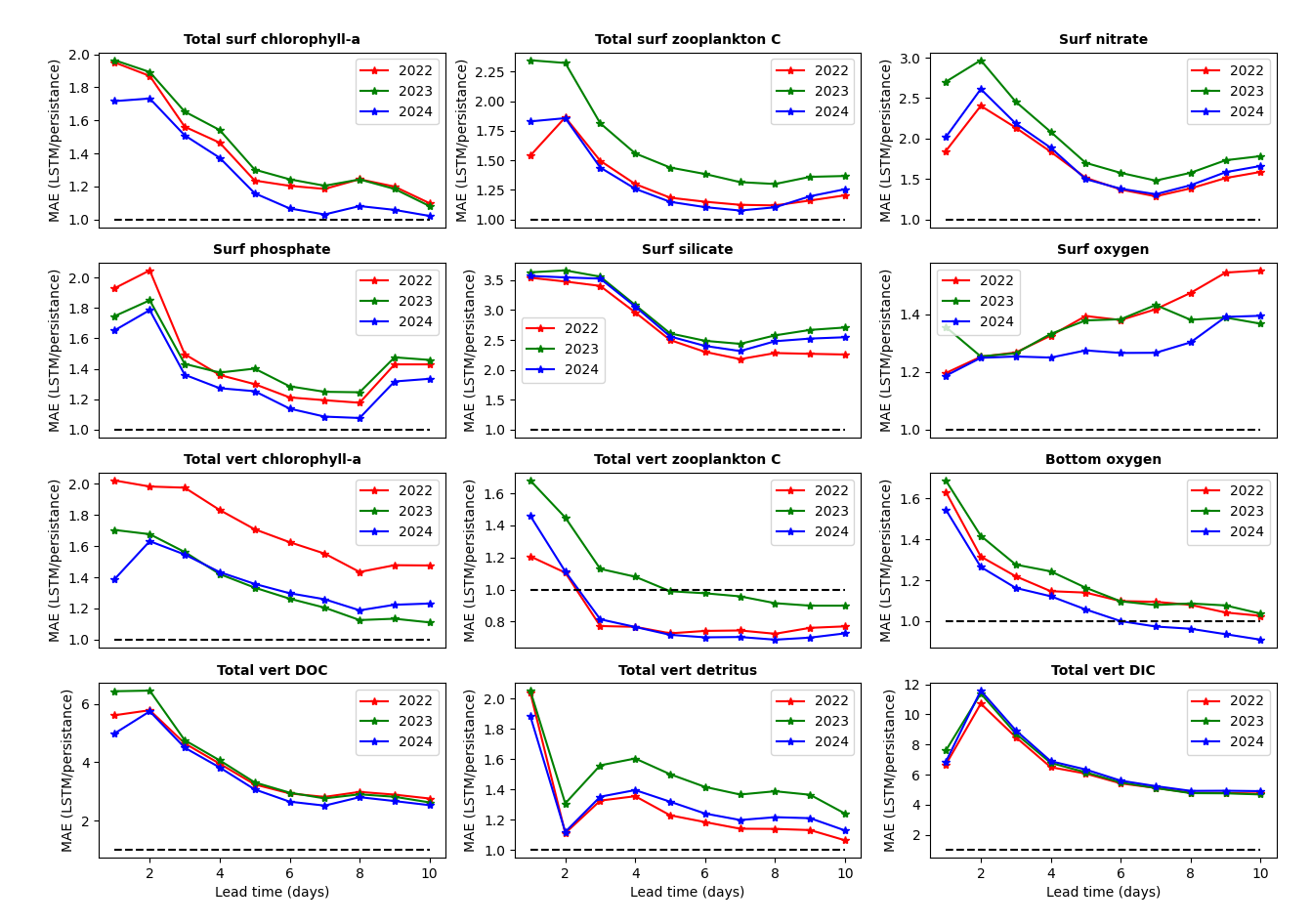}
    \caption{The RMSE skill of short-range (up to 10-day) prediction of 1D CNN emulator relative to the persistence.}
    \label{fig:A6}
\end{figure}

\begin{figure}[h!]
    \centering
    \includegraphics[width=1\linewidth]{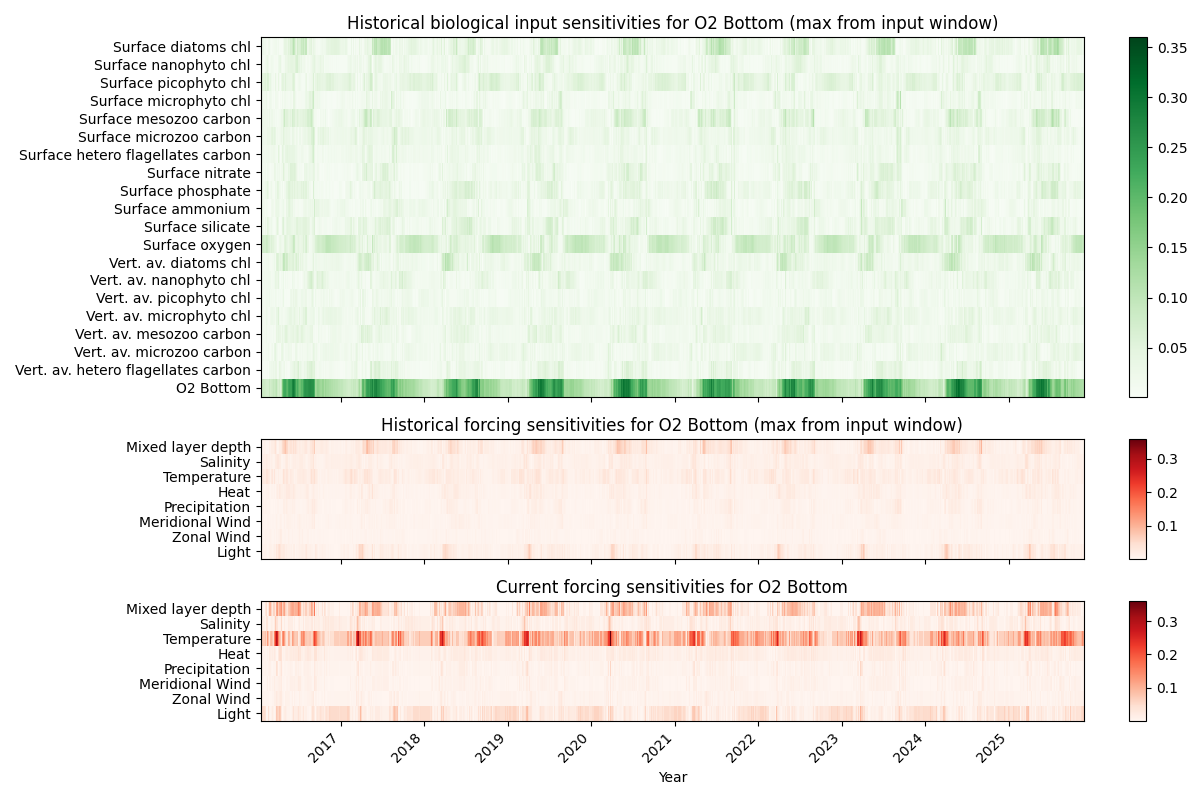}
    \caption{Temporal evolution of absolute sensitivity of oxygen bottom to each biological/forcing input, taking the maximum sensitivity within the 30-day lookback window for historic variables. Darker colours indicate periods of stronger sensitivity.}
    \label{fig:xai_hov_o2bottom}
\end{figure}

\end{document}